\documentclass[twocolumn]{aastex701}
\usepackage{amsmath}
\usepackage{graphicx}
\usepackage{indentfirst}
\usepackage{caption}
\usepackage{multirow}
\usepackage{longtable}
\usepackage{booktabs}
\usepackage{soul}

\usepackage{overpic}

\usepackage{floatrow}
\usepackage[labelfont=bf,labelformat=simple]{subfig}
\begin{document}

\title{On the Ultra-Long Gamma-Ray Transient GRB 250702B/EP250702a}

\correspondingauthor{Shao-Lin Xiong, Shu-Xu Yi}
\email{xiongsl@ihep.ac.cn, sxyi@ihep.ac.cn}

\author[0009-0007-6192-0213]{Jin-Peng Zhang}
\affil{State Key Laboratory of Particle Astrophysics, Institute of High Energy Physics, Chinese Academy of Sciences, Beijing 100049, China}
\affil{University of Chinese Academy of Sciences, Chinese Academy of Sciences, Beijing 100049, China}
\email{zhangjinpeng@ihep.ac.cn}

\author[0009-0008-8053-2985]{Chen-Wei Wang}
\affil{State Key Laboratory of Particle Astrophysics, Institute of High Energy Physics, Chinese Academy of Sciences, Beijing 100049, China}
\affil{University of Chinese Academy of Sciences, Chinese Academy of Sciences, Beijing 100049, China}
\email{cwwang@ihep.ac.cn}

\author[0009-0002-6411-8422]{Zheng-Hang Yu}
\affil{State Key Laboratory of Particle Astrophysics, Institute of High Energy Physics, Chinese Academy of Sciences, Beijing 100049, China}
\affil{University of Chinese Academy of Sciences, Chinese Academy of Sciences, Beijing 100049, China}
\email{zhyu@ihep.ac.cn}

\author[0000-0002-4771-7653]{Shao-Lin Xiong*}
\affil{State Key Laboratory of Particle Astrophysics, Institute of High Energy Physics, Chinese Academy of Sciences, Beijing 100049, China}
\email{xiongsl@ihep.ac.cn}

\author[0000-0001-7599-0174]{Shu-Xu Yi*}
\affil{State Key Laboratory of Particle Astrophysics, Institute of High Energy Physics, Chinese Academy of Sciences, Beijing 100049, China}
\email{sxyi@ihep.ac.cn}

\author[0009-0004-1887-4686]{Jia-Cong Liu}
\affil{State Key Laboratory of Particle Astrophysics, Institute of High Energy Physics, Chinese Academy of Sciences, Beijing 100049, China}
\affil{University of Chinese Academy of Sciences, Chinese Academy of Sciences, Beijing 100049, China}
\email{liujiacong@ihep.ac.cn}

\author[0000-0001-8664-5085]{Wang-Chen Xue}
\affil{State Key Laboratory of Particle Astrophysics, Institute of High Energy Physics, Chinese Academy of Sciences, Beijing 100049, China}
\affil{University of Chinese Academy of Sciences, Chinese Academy of Sciences, Beijing 100049, China}
\email{xuewangchen@ihep.ac.cn}

\author[0009-0006-5506-5970]{Wen-Jun Tan}
\affil{State Key Laboratory of Particle Astrophysics, Institute of High Energy Physics, Chinese Academy of Sciences, Beijing 100049, China}
\affil{University of Chinese Academy of Sciences, Chinese Academy of Sciences, Beijing 100049, China}
\email{tanwj@ihep.ac.cn}

\author[]{Zi-Rui Zhang}
\affil{College of Physics, Sichuan University, Chengdu, 610065, China}
\email{zhangzirui200410@163.com}

\author[0000-0001-8664-5085]{Rahim Moradi}
\affil{State Key Laboratory of Particle Astrophysics, Institute of High Energy Physics, Chinese Academy of Sciences, Beijing 100049, China}
\email{rmoradi@ihep.ac.cn}

\author{Hao-Xuan Guo}
\affil{State Key Laboratory of Particle Astrophysics, Institute of High Energy Physics, Chinese Academy of Sciences, Beijing 100049, China}
\affil{Department of Nuclear Science and Technology, School of Energy and Power Engineering, Xi'an Jiaotong University, Xi'an, China}
\email{guohx@ihep.ac.cn}

\author[0009-0001-7226-2355]{Chao Zheng}
\affil{State Key Laboratory of Particle Astrophysics, Institute of High Energy Physics, Chinese Academy of Sciences, Beijing 100049, China}
\affil{University of Chinese Academy of Sciences, Chinese Academy of Sciences, Beijing 100049, China}
\email{zhengchao97@ihep.ac.cn}

\author[0000-0001-5348-7033]{Yan-Qiu Zhang}
\affil{State Key Laboratory of Particle Astrophysics, Institute of High Energy Physics, Chinese Academy of Sciences, Beijing 100049, China}
\affil{University of Chinese Academy of Sciences, Chinese Academy of Sciences, Beijing 100049, China}
\email{yqzhang@ihep.ac.cn}

 \author[0009-0008-5068-3504]{Yue Wang}
 \affil{State Key Laboratory of Particle Astrophysics, Institute of High Energy Physics, Chinese Academy of Sciences, Beijing 100049, China}
 \affil{University of Chinese Academy of Sciences, Chinese Academy of Sciences, Beijing 100049, China}
\email{yuewang@ihep.ac.cn}
 
\author[0000-0001-9217-7070]{Sheng-Lun Xie}
\affil{State Key Laboratory of Particle Astrophysics, Institute of High Energy Physics, Chinese Academy of Sciences, Beijing 100049, China}
\affil{Institute of Astrophysics, Central China Normal University, Wuhan 430079, China}
\email{xiesl@ihep.ac.cn}

\author[0000-0002-8097-3616]{Peng Zhang}
\affil{State Key Laboratory of Particle Astrophysics, Institute of High Energy Physics, Chinese Academy of Sciences, Beijing 100049, China}
\affil{College of Electronic and Information Engineering, Tongji University, Shanghai 201804, China}
\email{zhangp97@ihep.ac.cn}

\author{Yang-Zhao Ren}
\affil{School of Physical Science and Technology, Southwest Jiaotong University, Chengdu Sichuan, 611756, China}
\affil{State Key Laboratory of Particle Astrophysics, Institute of High Energy Physics, Chinese Academy of Sciences, Beijing 100049, China}
\email{renyz@ihep.ac.cn}  

\author[0000-0001-5798-4491]{Cheng-Kui Li}
\affil{State Key Laboratory of Particle Astrophysics, Institute of High Energy Physics, Chinese Academy of Sciences, Beijing 100049, China}
\email{lick@ihep.ac.cn}

\author[0000-0003-4585-589X]{Xiao-Bo Li}
\affil{State Key Laboratory of Particle Astrophysics, Institute of High Energy Physics, Chinese Academy of Sciences, Beijing 100049, China}
\email{lixb@ihep.ac.cn}

\author[0000-0002-6540-2372]{Ce Cai}
\affil{College of Physics and Hebei Key Laboratory of Photophysics Research and Application, Hebei Normal University, Shijiazhuang, Hebei 050024, China}
\email{caice@hebtu.edu.cn}

\author[0000-0003-2957-2806]{Shuo Xiao}
\affil{Guizhou Provincial Key Laboratory of Radio Astronomy and Data Processing, 
 Guizhou Normal University, Guiyang 550001, China}
\affil{School of Physics and Electronic Science, Guizhou Normal University, Guiyang 550001, China}
\email{xiaoshuo@gznu.edu.cn}

\author[0000-0003-0274-3396]{Li-Ming Song}
\affil{State Key Laboratory of Particle Astrophysics, Institute of High Energy Physics, Chinese Academy of Sciences, Beijing 100049, China}
\affil{University of Chinese Academy of Sciences, Chinese Academy of Sciences, Beijing 100049, China}
\email{songlm@ihep.ac.cn}

\author[0000-0001-5586-1017]{Shuang-Nan Zhang}
\affil{State Key Laboratory of Particle Astrophysics, Institute of High Energy Physics, Chinese Academy of Sciences, Beijing 100049, China}
\affil{University of Chinese Academy of Sciences, Chinese Academy of Sciences, Beijing 100049, China}
\email{zhangsn@ihep.ac.cn}

\begin{abstract}

GRB 250702B/EP250702a is an interesting long-duration gamma-ray transient whose nature is in debate.
To obtain a full picture in gamma-ray band, we implement a comprehensive targeted search of burst emission in a wide window of 30 days jointly with \textit{Insight}-HXMT, GECAM and \textit{Fermi}/GBM data
within the ETJASMIN framework. 
In gamma-ray band, we find there is a 50-second precursor about 25 hours before the 4-hour main burst, which generally consists of 4 emission episodes.
Remarkably, we find that the soft X-ray emission after the main burst 
decays as a power-law with start time aligning with the last episode of main emission and index of -5/3 which is perfectly consistent with the canonical prediction of fallback accretion. We conclude that the properties of precursor, main burst and the following soft X-ray emission strongly support the Ultra-Long Gamma-Ray Burst (ULGRB) scenario
and all these gamma-ray and soft X-ray emission probably originate from relativistic jet whose luminosity is dominated by the fallback accretion rate during the death collapse of a supergiant star. 

\end{abstract}


\section{Introduction}

As the most violent phenomenon in the universe, Gamma-Ray Bursts (GRBs) manifest themselves as bright gamma-ray emission from distant galaxies. 
GRBs can be traditionally classified into long GRB and short GRB based on the duration of gamma-ray emission \citep[e.g.][]{GRB_T90_classification}. 
Long GRBs are usually resulted from the core collapse of massive star (Type II GRB) \citep{Woosleycollapsar1993,Paczycollapsar1998,Woosleycollapsar2006}, 
while short GRBs are generally believed to originate from merger of compact binaries (Type I GRB) \citep{Blinnikovmerger1984,Paczynskimerger1986,1Meszarosmerger1992,Limerger1998,GWEM_170817A}. 

However, 
an increasing number of GRBs are reported to deviate the ``long-collapsar, short-merger" classification pattern, including long duration GRBs with merger origin (e.g. GRB 060614, GRB 211211A, GRB 211227A, GRB 230307A) \citep{060614,YJ_GRB211211A_nature,21227A,2025NSRSun} and short duration GRBs with collapsar origin (e.g. GRB 200826A, GRB 230812B, GRB 240825A) \citep{26A_zhang_2021,12B_Wang_2024,25A_wang_2025}. 
Furthermore, some merger GRBs (including GRB 211211A and GRB 230307A) feature an intrinsically long main burst and thus are suggested to form a special sub-class of Type I GRB, i.e. Type IL GRB \citep{TypeIL_1,TypeIL_2,TypeIL_3}. How the compact merger system can produce long duration burst is an open question.

Similar with the rarity of long-duration Type I GRBs, very long duration Type II GRBs are also scarce:
only a very small fraction of Type II GRBs can last more than thousands of seconds, named as Ultra-Long GRB (ULGRB). Whether ULGRB just represents a tail at the long-end of the distribution of Type II GRB, or belongs to a distinct new class with a different physical origin, is inconclusive yet \citep{virgili_grb_2013, levan_new_2013, boer_are_2015}.
There are suggestions that ULGRBs are produced by supergiant star collapse \citep{Supergiant_1,Supergiant_2}, GRBs with central engine of magnetar \citep{greiner_very_2015}, or IMBH-WD tidal disruption \citep{levan_new_2013}.

Remarkably, some high-energy transients with extremely long duration initially considered to be GRB or ULGRB are finally identified as jetted Tidal Disruption Event (TDE), such as Swift J1644+57 \citep{J1644}. 
In general consideration, TDEs are produced by the disruption of a star by a dormant supermassive black hole when the star passes too close to the black hole, or a White dwarf (WD) by an intermediate-mass black hole \citep{1982_TDE_galaxy,1988Natur.333..523R}.
In some cases, a relativistic jet could be launched, leading to jetted TDE. 
Recently, more jetted TDE candidates have been found, such as Swift J2058+05 \citep{J2058}, and AT2022cmc \citep{AT2022cmc}. 
Determining whether a very long high energy transient is a jetted TDE or
a ULGRB is helpful to probe the nature of the transient. Unfortunately, this is usually a very challenging task because of the rare occurrence and fast-fading nature of these astronomical transient events.

Recently, a peculiar long-duration high energy transient is discovered, denoted as GRB 250702B/EP2050702a, which is first reported by \textit{Fermi}/GBM as a series of individual GRBs: GRB 250702B, GRB 250702D and GRB 250702E \citep{GBM_GCN_250702B,GBM_GCN_250702D,GBM_GCN_250702E,GBM_GCN_250702BDE} on July 2nd, 2025.
The gamma-ray\footnote{In this work, gamma-ray means photons with energy greater than about 10 keV, which is also called soft gamma-ray or hard x-ray. On contrast, soft x-ray photons with energy below about 10 keV.} emission of this event is also detected by Konus-Wind \citep{KW_GCN}, SVOM/GRM \citep{GRM_GCN}, Swift/BAT \citep{BAT_GCN}, \textit{Insight}-HXMT and GECAM-B. 
Prompt soft X-ray emission is detected by EP/WXT \citep{EP_GCN_1} and MAXI \citep{MAXI_GCN}.
Interestingly, stacking data of EP/WXT suggest that this transient has already emerged in soft X-ray band about one day before the gamma-ray detection of GRB 250702B \citep{EP_GCN_1}. 
X-ray follow-up observations with Swift/XRT \citep{XRT_GCN}, EP/FXT \citep{EP_GCN_2} and hard X-ray of NuSTAR \citep{NUSTAR_GCN} reveal a fading X-ray source with a featureless absorbed power-law spectrum.
Observations in optical, near-IR and radio are also preformed. Especially, \citet{Levan_GCN} reported an extremely red fast-fading counterpart located on the non-nuclear region of a disc-like galaxy with HST. 
The EP location of this source is RA=284.6895\,deg, Dec=-7.8738\,deg \citep{EP_GCN_1}. Redshift ($z$) of this source is measured to be 1.036 \citep{gompertz_jwst_2025}.

Despite of many observations, the origin of GRB 250702B/EP250702a is on hot debate \citep[e.g.][]{2025arXiv250714286L, beniamini_ultra-long_2025, carney_opticalinfrared_2025, oconnor_comprehensive_2025, eyles-ferris_can_2025, gompertz_jwst_2025, li_fast_2025, neights_grb_2025, oganesyan_ultra-long_2025, song_possible_2025, an_precessing_2025, granot_milli-tidal_2025}. Especially, possibilities of ULGRB or jetted TDE have been proposed but cannot be decisively determined \citep[e.g.][]{2025arXiv250714286L}. In this work, we primarily investigate the high energy emission (including gamma-ray and soft X-ray emission) of this source and obtain various interesting discoveries from several episodes of this transient --- pre-burst, main burst gamma-ray activity, and post-burst X-ray emission.

This paper is organized as follows.
Search for burst components and analyses covering the full burst episodes in gamma-ray band with \textit{Insight}-HXMT/HE, GECAM-B and \textit{Fermi}/GBM are described in Section~\ref{section2}. 
Temporal analysis of soft X-ray emission detected by Swift/XRT and Chandra are presented in Section~\ref{section3}, focusing on the power-law decay slope with different start time. 
Discussions on the plausible progenitor models are depicted in Section~\ref{section4}, followed by interpretation of the observations in Section \ref{sec_discussions}. Summary and conclusion are given in section~\ref{section5}.

Please note that all parameter errors in this work are for 68\% confidence level if not otherwise stated.

\section{Gamma-ray emission}\label{section2}

\subsection{Search for burst component}

For this source, there are several triggers, initially named as GRB 250702B, GRB 250702D and GRB 250702E, reported by Fermi/GBM \citep{GBM_GCN_250702B}, SVOM/GRM \citep{GRM_GCN}, etc. We point out that the monitoring coverage to the source by individual instrument is usually incomplete due to the Earth blocking or instrument turn-off flying over the SAA region in orbit. Konus-Wind may have a complete time coverage, but its reported light curve appears to be limited to about 5 hours \citep{KW_GCN}. Whether there is other burst component outside these reported time windows, possibly separated by hours or even days, is unclear.

\begin{figure*}
\centering
\begin{tabular}{cc}
    \hspace{22pt}
    \begin{overpic}[width=0.4\textwidth]{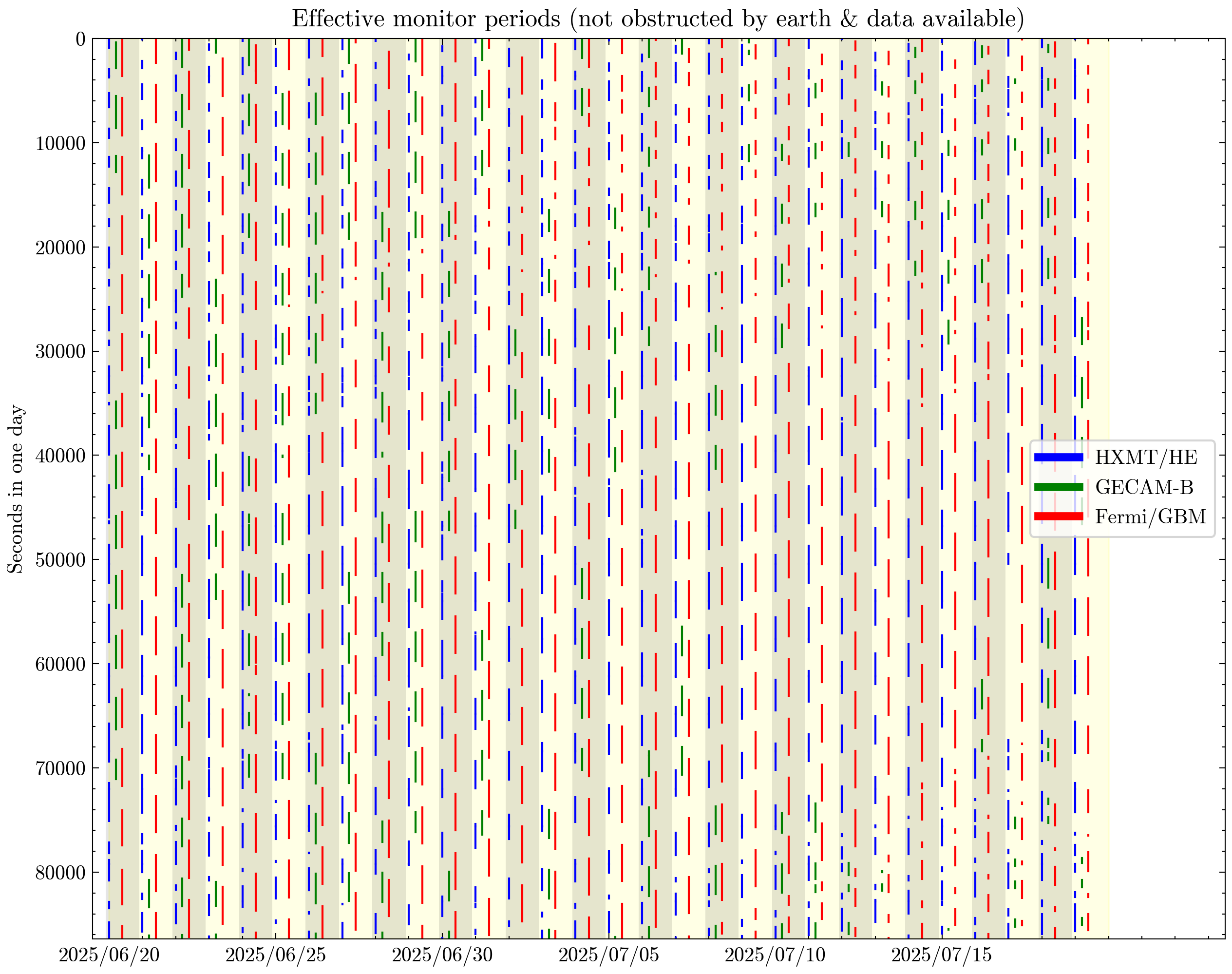}\put(0, 74){\bf a}\end{overpic} &
    \begin{overpic}[width=0.42\textwidth]{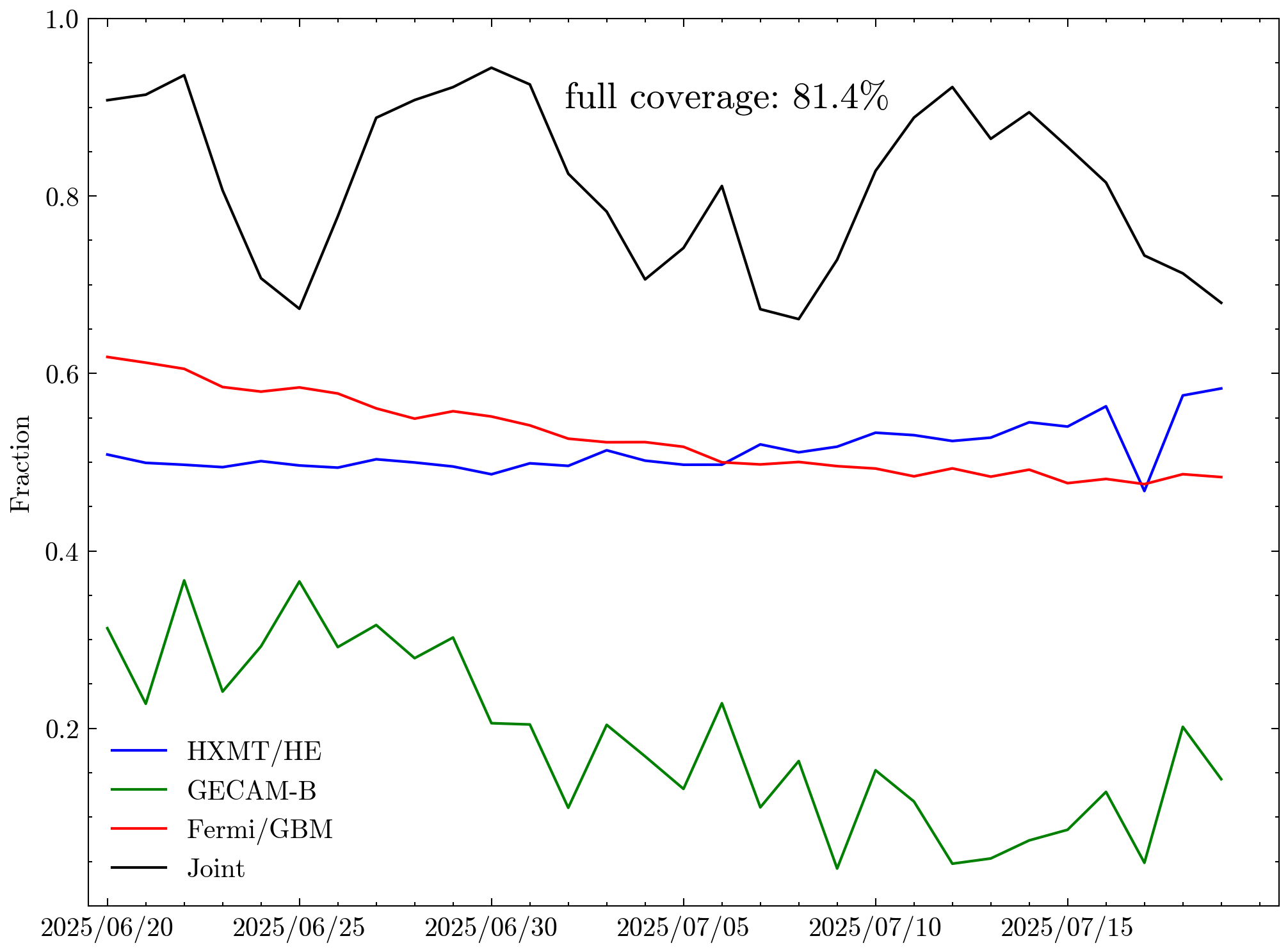}\put(0, 70){\bf b}\end{overpic} \\
    \multicolumn{2}{c}{\begin{overpic}[width=0.9\textwidth]{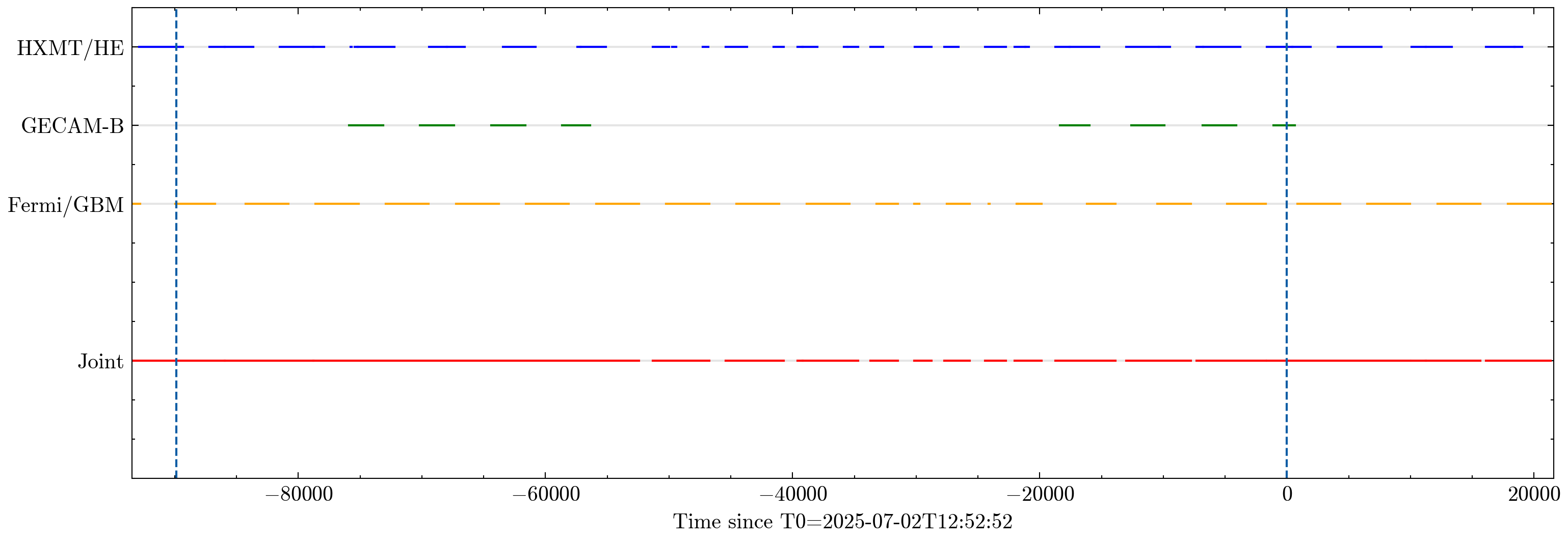}\put(-2, 33){\bf c}\end{overpic}} \\
\end{tabular}
\caption{ \textbf{Panel (a):} Monitoring periods for each instrument: \textit{Insight}-HXMT/HE, GECAM-B or \textit{Fermi}/GBM. Each column represents one day. Time goes by from up to down in each column. \textbf{Panel (b):} Coverage percentage for each instrument or joint observations. The joint monitored time intervals cover 81.4\% of the 30-day search. \textbf{Panel (c):} Monitored time intervals from precursor to end of main burst stage for each instrument or joint observations.}
\label{fig:cover}
\end{figure*}

In attempt to provide a better view of this source with emphasis on both the wide time coverage and the high detection sensitivity, we performed a comprehensive targeted coherent search from about -10 days to +20 days relative to GRB 250702B, using the ETJASMIN pipeline \citep{cai_energetic_2025,Cai_search_HXMT} with \textit{Insight}-HXMT/HE, GECAM-B and \textit{Fermi}/GBM data. At each time interval, only those instruments that are visible to the source (not blocked by the Earth) and have available data are used for the burst search (see Fig\,\ref{fig:cover}).

In the search, background is derived from a recursively fitting with 4th-order polynomial on 1\,s binned light curve based on goodness-of-fit. This method makes background estimation more robust, preventing potential false triggers caused by the in-and-out-of-SAA background variation or cosmic X-ray source occultation. The significance calculation method proposed in \cite{blackburn_high-energy_2015} is adopted in this search. The search time scale is from 0.1\,s to 4\,s. We note that our search focus on the fast variation emission.
However, slow-varying emission components accompanying the fast component could be also identified in the light curve of the burst if they indeed exist and are bright enough. 

\subsection{Burst episodes}

According to our targeted search of this source, we find there is a relatively short emission around 2025-07-01 11:55 UTC and a series of much longer burst emissions
from about 2025-07-02 12:50 to 16:30 UTC, as shown in Fig.~\ref{fig:lc_overview}.
There is no other significant burst component found in gamma-ray band (20\,keV to 3\,MeV) from this source from -10 days to +20 days relative to GRB 250702B.
Seeing from the morphology in the light curve and following the convention of GRBs, we indicate the burst emission at 2025-07-01 as precursor (denoted as P) and that at 2025-07-02 as main burst (denoted as M, including M1 to M4). Detailed light curves in higher temporal resolution of all these episodes (P, M1 to M4) are shown in Fig.~\ref{fig:precursor} to \ref{fig:M4}.

We notice that the joint light curve of \textit{Insight}-HXMT/HE, GECAM-B and \textit{Fermi}/GBM of the main burst is generally consistent with that of Konus-Wind \citep{KW_GCN}.
However, EP/WXT detected soft X-ray emission from 02:53:44 UTC to 22:32:44 UTC on July 2nd \citep{EP_GCN_1} which is much longer and well embraces the main burst episode in gamma-ray band. In addition, EP/WXT also detected signal from this burst on July 1st with stacking analysis \citep{EP_GCN_1}, which lends additional support to our discovery of the precursor. We point out that this feature that soft X-ray emission spans wider in time than that of gamma-ray emission is often seen in GRBs \cite[e.g.][]{gao_soft_2025, zhang_multi-instrument_2025}.

\subsubsection{Main burst}\label{section2.2.1}

We find that the main burst can be generally divided into 4 episodes (denoted as M1 to M4) as they are well separated by the quiescent regions in the light curves (Fig. \ref{fig:lc_overview_all}). We note that, by ``quiescent'' we mean the emission is either absent or under the sensitivity of the used instruments in this work, i.e. \textit{Insight}-HXMT, GECAM-B and \textit{Fermi}/GBM.

Interestingly,
for the first episode (M1), \textit{Insight}-HXMT and GECAM-B detected the early burst emission starting from about 2025-07-02T12:52:52 UTC (denoted as $T_0$ in this work) which is well before the \textit{Fermi}/GBM trigger time (13:09:02 UTC) of this event (literally used to name as GRB 250702D), as a large part of the emission in episode M1 is invisible to \textit{Fermi}/GBM due to the Earth occultation, as shown in Fig. \ref{fig:M1}.
Three pulses are clearly visible in the \textit{Insight}-HXMT/HE light curve from $T_0$-50 to +100\,s in Fig. \ref{fig:M1}.
\textit{Insight}-HXMT/HE also detected many pulses in the time region where \textit{Fermi}/GBM is occulted by the Earth.
Some of the \textit{Insight}-HXMT/HE pulses are also detected by GECAM-B, confirming these early emission. 

Detailed light curves for four episodes of the main burst are shown in Fig. \ref{fig:M1} -- \ref{fig:M4}. Each episode consists of multiple fast pulses and some underlying slow emissions. The temporal structures in each episode are generally consistent between \textit{Insight}-HXMT/HE and \textit{Fermi}/GBM. The underlying emissions are more obvious in \textit{Fermi}/GBM light curves, which covers a lower energy band than \textit{Insight}-HXMT/HE. 

According to the background-subtracted and normalized light curves of \textit{Insight}-HXMT/HE in higher energy range (200\,keV--3\,MeV) and \textit{Fermi}/GBM and GECAM-B including a lower energy range (50\,keV - 2\,MeV), as shown in Fig. \ref{fig:lc_overview_all}, the first episode (M1) appears to be relatively harder in spectrum than the later three episodes (M2 to M4) of main burst. M1 also seems to have the longest duration among of all episodes. Besides, we find that there is no significant trend of monotonic increase or decrease in the fluence from episode M1 to M4.

Among the whole observation interval of main burst episode, we confirmed some other signals in \textit{Fermi}/GBM data. There is one short burst (GRB 250702C, \citet{neights_fermi_2025}) overlapped on the M3 episode, and one solar flare at the M4 episode. The short burst lasts only for less than a few seconds. The solar flare comes from a nearly opposite direction with respect to the GRB 250702B, thus won't affect our analysis with selected detectors.

\subsubsection{Precursor}

\begin{figure*}
    \centering
    \begin{minipage}{.8\linewidth}
        \centering
        \sidesubfloat[]{
        \includegraphics[width=.9\linewidth]{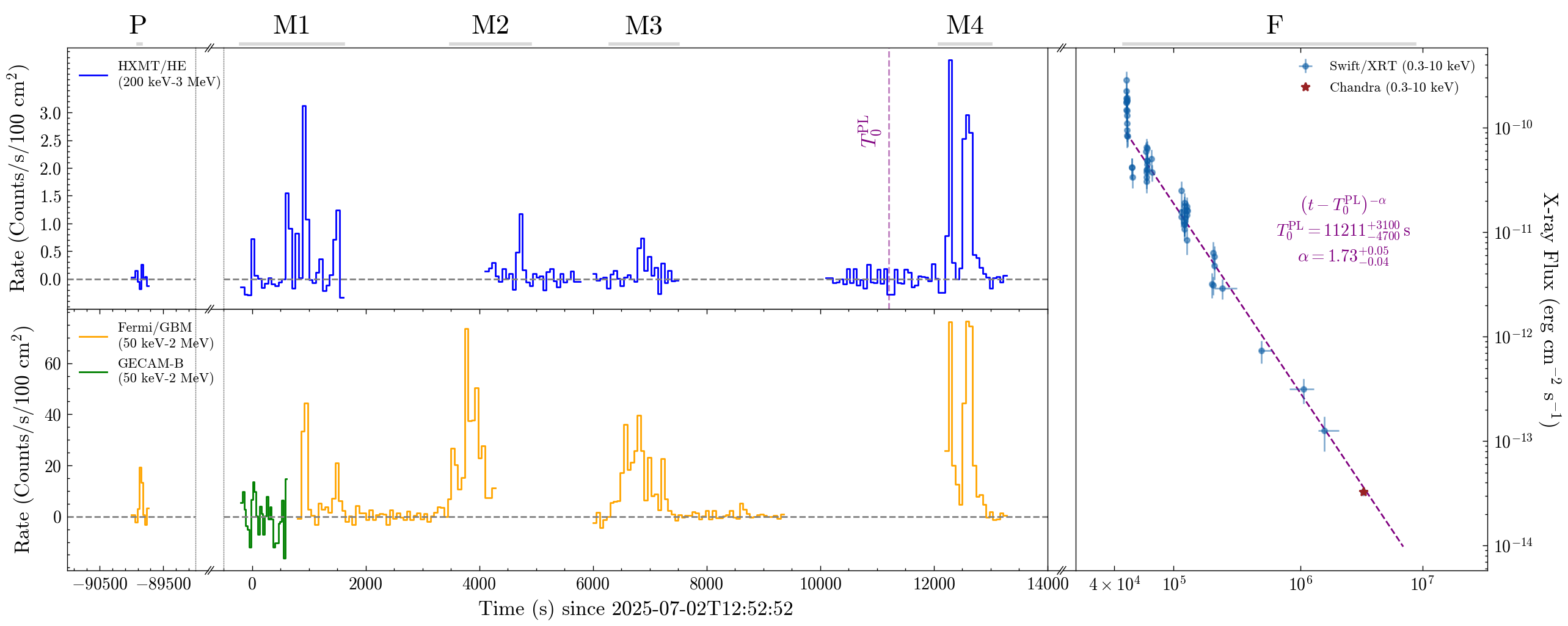}
        \label{fig:lc_overview_all}
        }
        \\
        \begin{minipage}[t]{.49\linewidth}
            \centering
            \begin{tabular}{c}
                \sidesubfloat[]{
                \includegraphics[width=.8\linewidth]{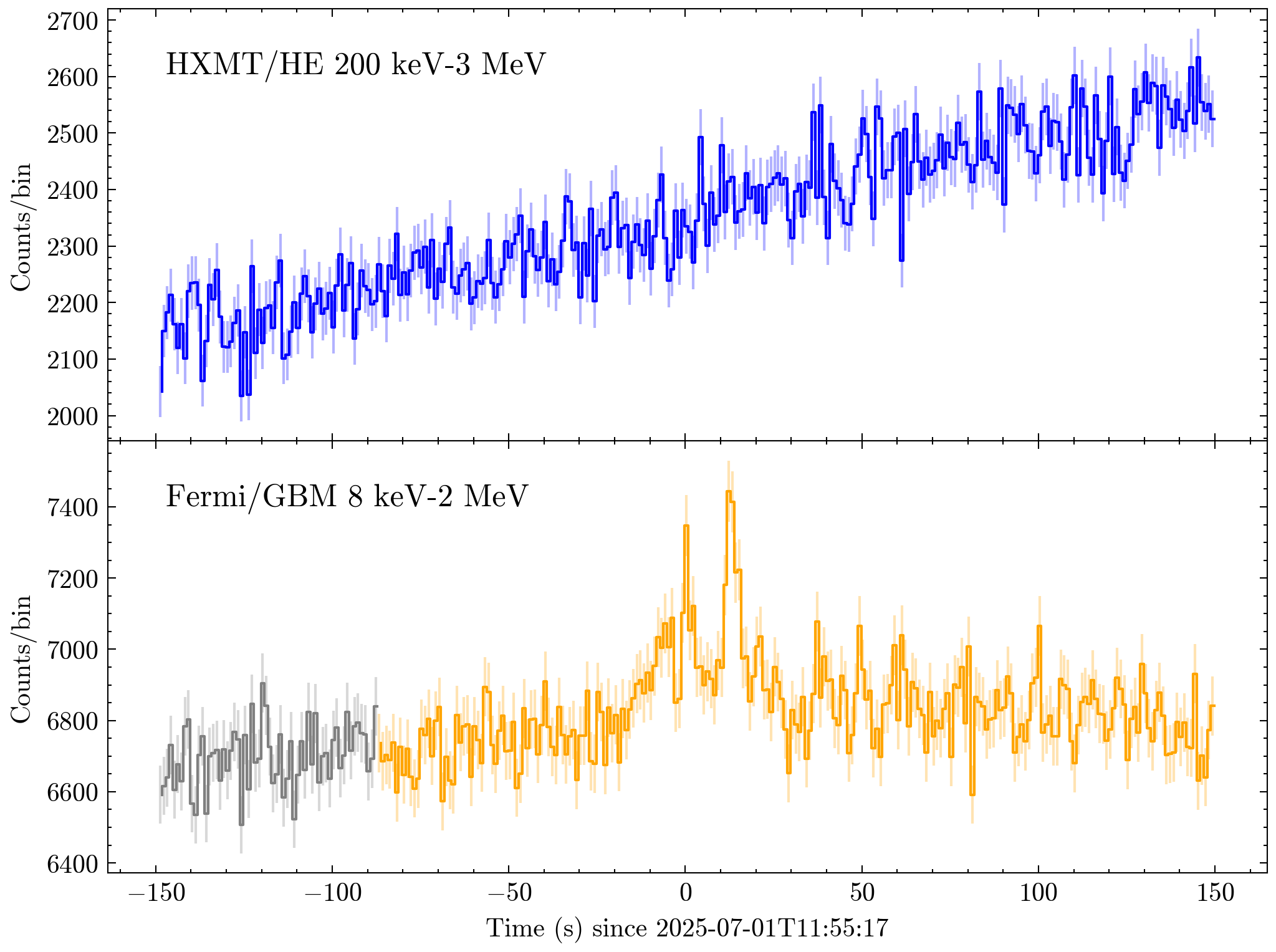}
                \label{fig:precursor}
                }
                \\
                \sidesubfloat[]{
                \includegraphics[width=.8\linewidth]{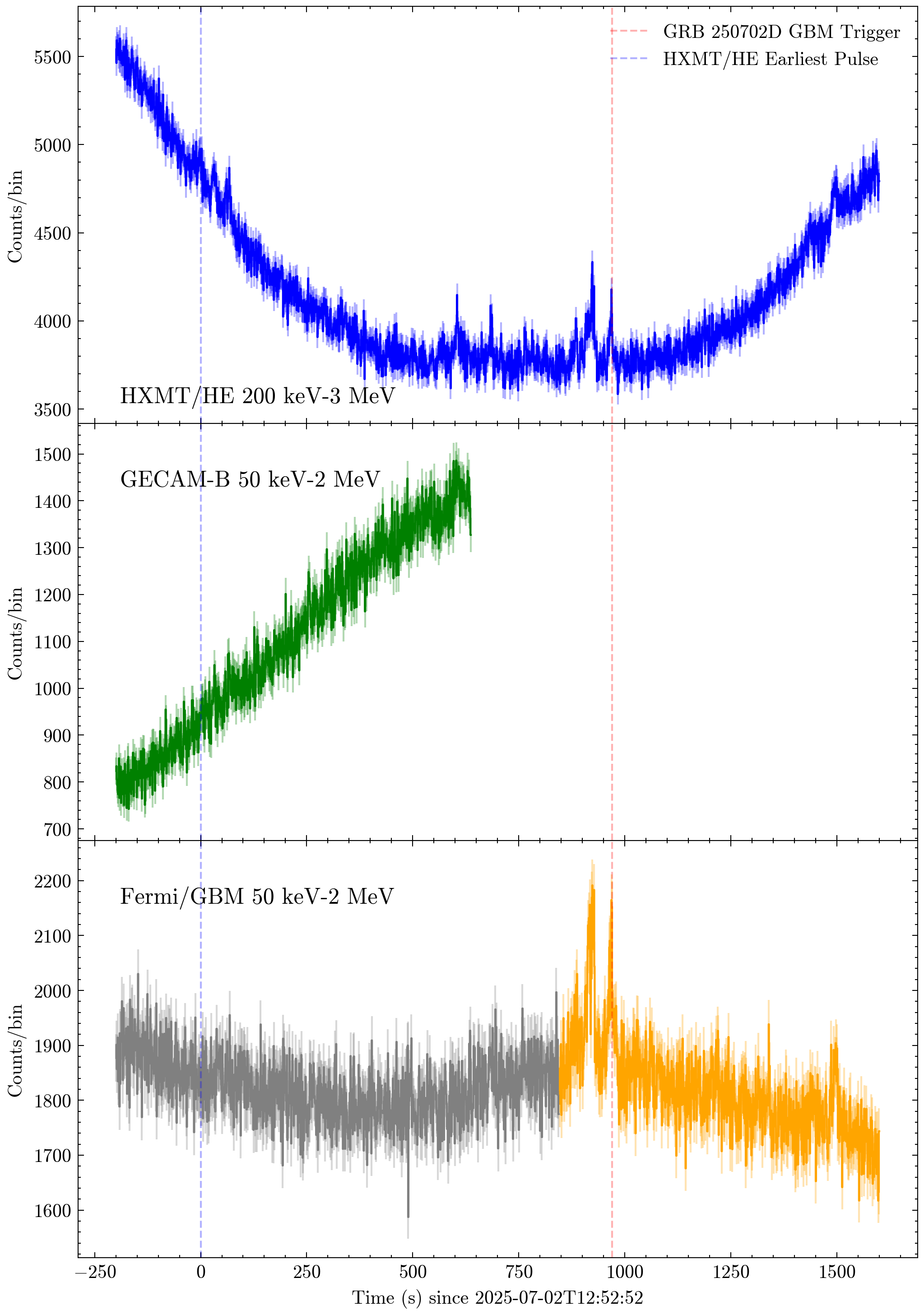}
                \label{fig:M1}
                }
            \end{tabular}
        \end{minipage}
        \begin{minipage}[t]{.49\linewidth}
            \centering
            \begin{tabular}{c}
                \sidesubfloat[]{
                \includegraphics[width=.8\linewidth]{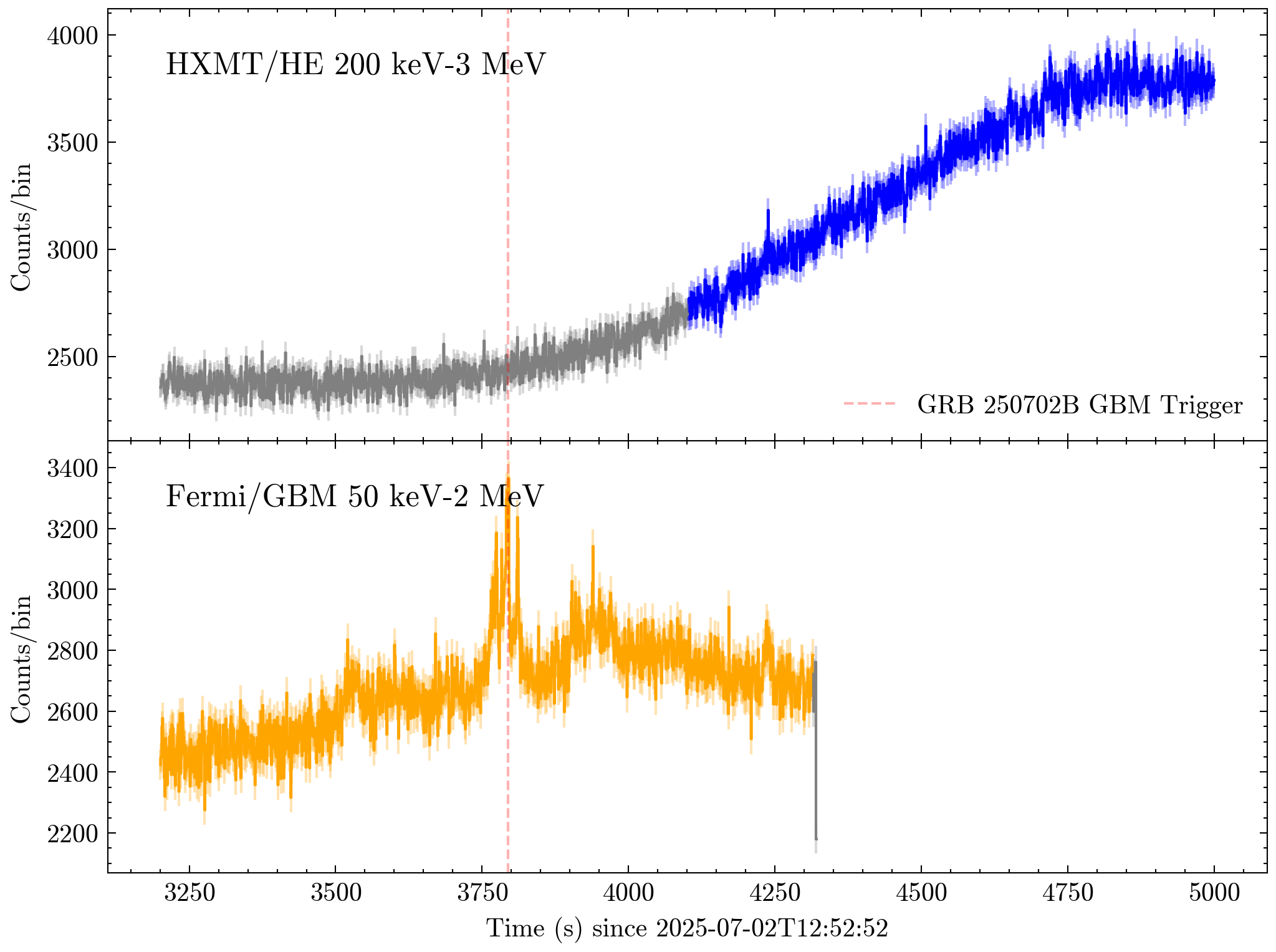}
                \label{fig:M2}
                }
                \\
                \sidesubfloat[]{
                \includegraphics[width=.8\linewidth]{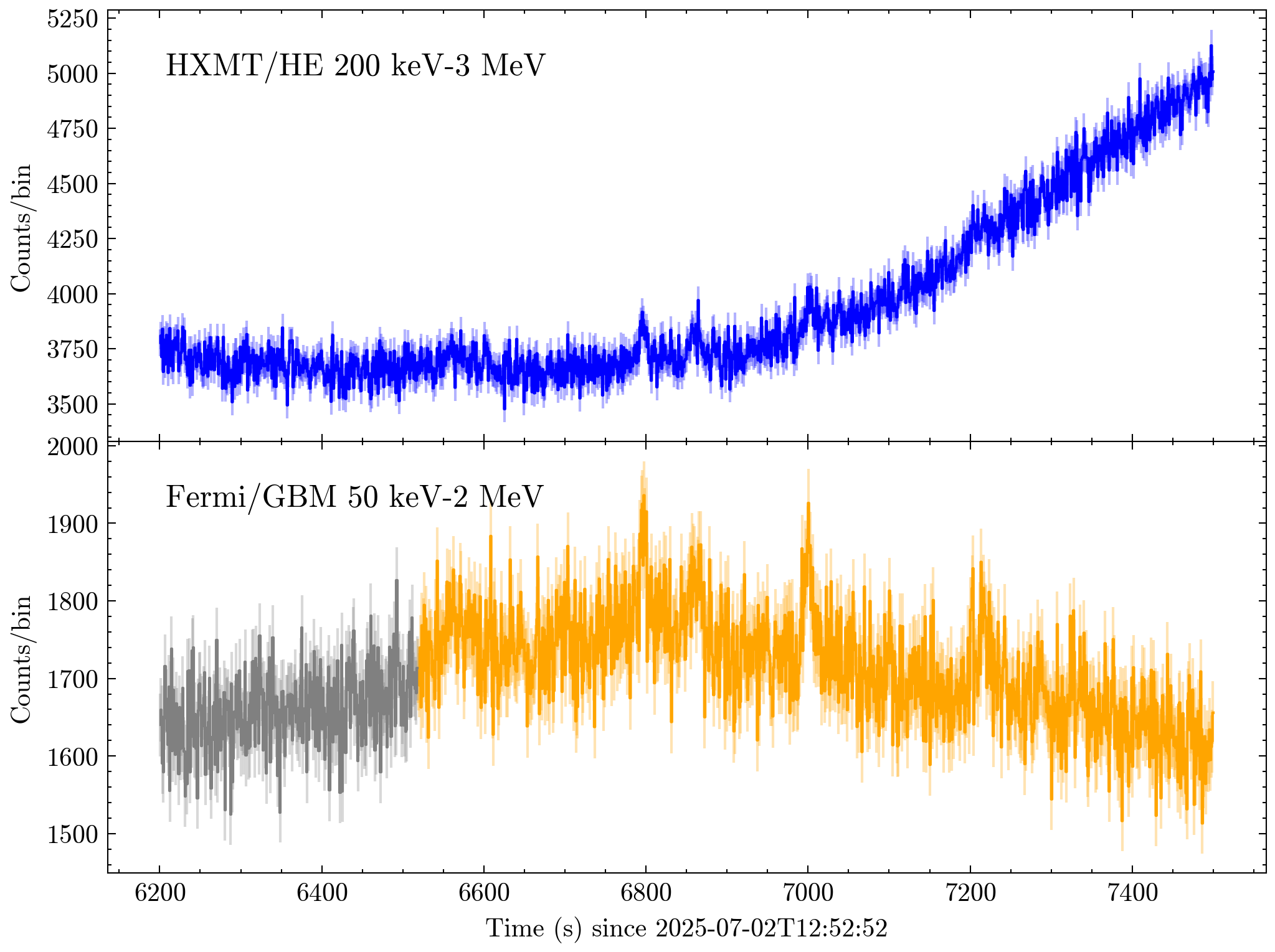}
                \label{fig:M3}
                }
                \\
                \sidesubfloat[]{
                \includegraphics[width=.8\linewidth]{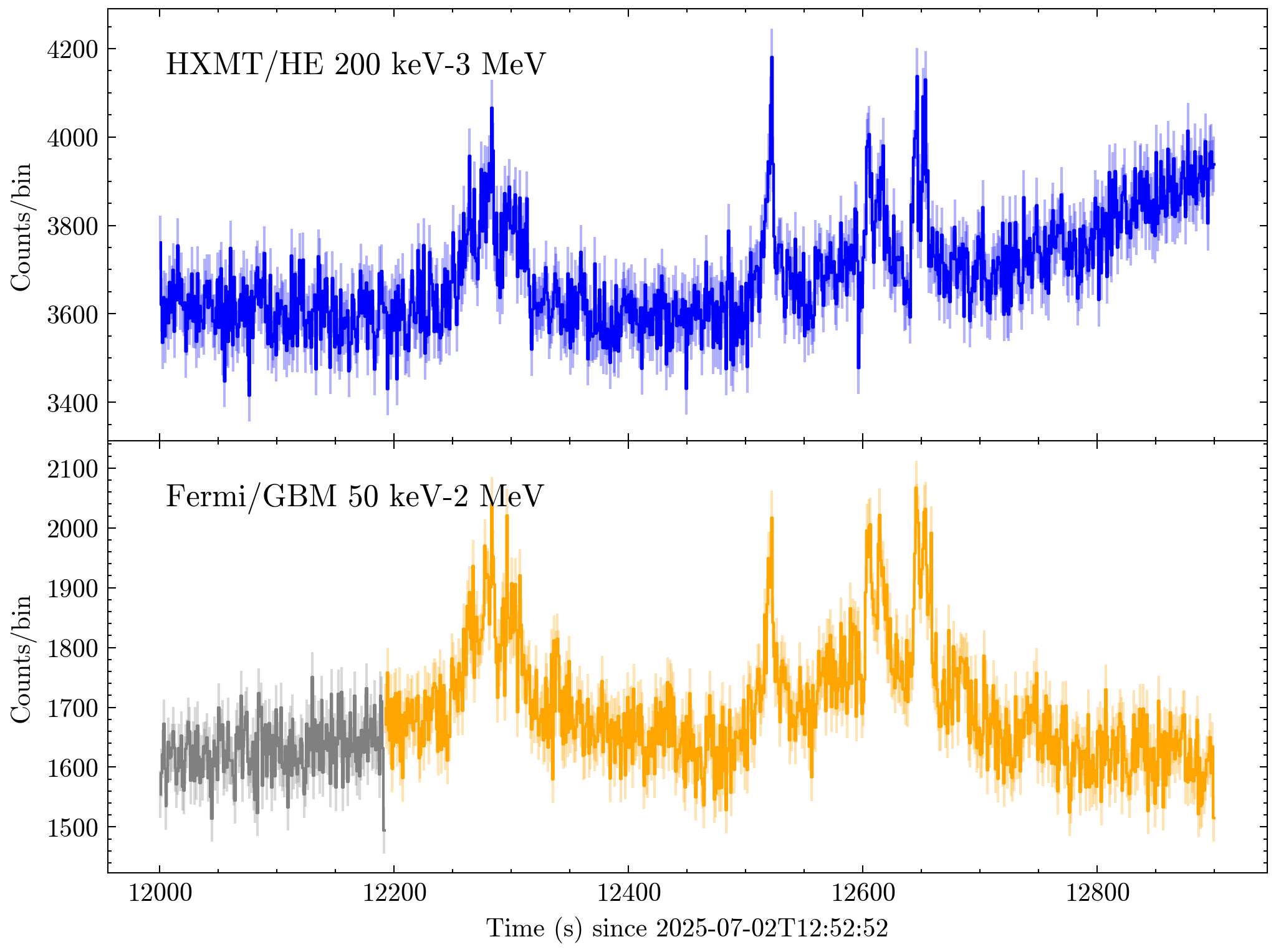}
                \label{fig:M4}
                }
            \end{tabular}
        \end{minipage}
    \end{minipage}

    \caption{\textbf{Panel (a):} Gamma-ray and soft X-ray light curves of GRB 250702B/EP250702a in a wide time window of about 40 days.
    Gamma-ray light curves are background subtracted and normalized by receiving area of detectors, from \textit{Insight}-HXMT/HE, GECAM-B and \textit{Fermi}/GBM. Precursor light curves of \textit{Insight}-HXMT/HE and \textit{Fermi}/GBM, and main burst light curve of GECAM-B are binned in 30 seconds, while others in 60 seconds. Soft X-ray data are well fitted with a power-law with index of around -5/3. Details of data analyses could be found in text and appendix.
   \textbf{Panel (b) - (f):} Light curves of precursor (P) and four episodes (M1 to M4) of main burst, binned in 1 second.}
   \label{fig:lc_overview}
\end{figure*}

Surprisingly, we find
that, owing to the sensitive targeted coherent search, there is a significant burst in GBM data which triggered our search program at 2025-07-01T11:55:17 UTC, which is about 25 hours before the start time of main burst. Localization of this burst with \textit{Fermi}/GBM GDT package \citep{GDT-Core} shows 
that the direction of GRB 250702B/EP250702a falls within the 3-$\sigma$ region of the skymap.
The localization region contains several known magnetars, but considering the total duration of about 50 seconds of this burst, which can be seen in Fig. \ref{fig:precursor}, it is quite unusual in magnetar bursts duration distribution \citep[e.g.][]{lin_detailed_2013, xie_gecam_2025}. We note that there is no other signal (except for known GRBs) with over 1\,s duration triggered our search program with a higher significance than this one in the 1-month coherent targeted search, indicating that this signal is not accidentally coincident with this burst. We also note that a subthreshold detection of GRB with similar flux and fluence as this signal in Fermi/GBM data is not uncommon, such as GRB 251021A \citep{ravasio_grb_2025}. Further taking EP/WXT detection around this signal into account \citep{EP_GCN_1}, we conclude that this signal is very likely a part of GRB 250702B. 
No gamma-ray transient emission is found between this signal and the main burst, which is the reason we call it precursor, following the convention of GRB.

This precursor primarily consists of several pulse, with a total duration about 50 seconds. We note that the joint observation of \textit{Insight}-HXMT and \textit{Fermi}/GBM provides a complete monitoring of a much wider time window embracing this burst but found no other burst components (Fig.~\ref{fig:precursor}). Therefore, we conclude that this precursor only lasts for about 50 seconds, much shorter than the episodes in the main burst. However, we notice that the precursor exhibits a minimum time variability timescale compatible to that of main burst.

Non-detection of this precursor by \textit{Insight}-HXMT/HE indicates that its spectrum is relatively soft. Indeed, there is no signal above 150\,keV in GBM data, well compatible with the non-detection of \textit{Insight}-HXMT/HE. We note that EP/WXT did not observe this source at the time of the precursor.

Two most significant detectors, n8 and nb, are selected for spectral fit. To estimate the background count rate at each PHA channel, we use 4-th order polynomial to fit the 10 s binned light curve for each channel in certain time intervals (-130 to -30 s, and 50 to 150 s). Total count spectrum, in 127 channels (among all 128 channels, the last channel is overflow channel and thus is excluded), is calculated from -10 to +20 s and subtracted by background count spectrum to get the net count spectrum.

We use \texttt{ELISA} \citep{ELISA} to do spectral analysis for this burst. \texttt{ELISA} is a Bayesian posterior estimation program, which serves as a canonical way to solve inaccuracy induced by simple assumption in Likelihood/chi square method \citep[e.g.][]{protassov_statistics_2002}.

The spectral analysis of the GBM data shows that this precursor can be adequately fit with power-law model with index of $1.65^{+0.10}_{-0.10}$.
The fluence of the precursor is $3.4_{-0.5}^{+0.6}\times 10^{-6}\, \rm erg / cm^2$, and the 1-second peak flux is $3.2_{-0.7}^{+0.8} \times 10^{-7} \rm erg / cm^2 / s$ (8--1000\,keV). As an estimation in order of magnitude, the main burst has a total fluence of about $10^{-3} \rm erg / cm^2$ (see also \citet{KW_GCN, neights_fermi_2025}).
Considering the redshift \citep{gompertz_jwst_2025}, the isotropic energy of the precursor and main burst is $1.86_{-0.29}^{+0.31} \times 10^{52} \rm erg$ and  $\sim 10^{55} \rm erg$, respectively.

\subsubsection{Relation between Waiting time and duration}

\begin{figure}
    \centering
    \includegraphics[width=\linewidth]{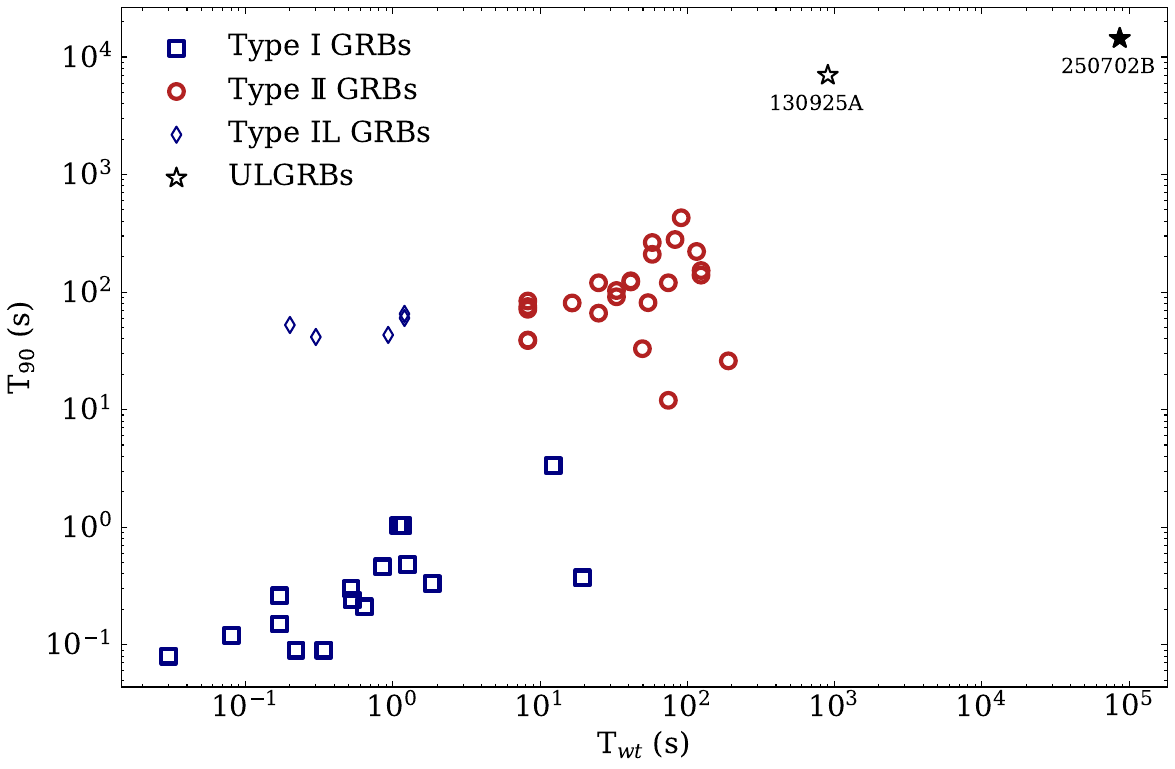}
    \caption{Relation between precursor waiting time and main burst duration. 
    }
    \label{fig:twt-T90}
\end{figure}

The waiting time ($T_{\rm WT}$) between precursor and main burst is about 25 hours, while the duration of the main burst ($T_{90}$) is about 4 hours. We compared this burst to all kinds of GRBs in the $T_{\rm WT}$ - $T_{90}$ diagram and find this burst is generally compatible to the trend of GRBs, as shown in Fig \ref{fig:twt-T90}.

Note that there is reported soft X-ray emission by EP/WXT during the apparent quiescent time between precursor and main burst in gamma-ray band. The fact that there is ongoing (soft X-ray) emission during the so-called quiescent time is not unexpected but well consistent with the picture demonstrated in e.g. GRB 230307A \citep{TypeIL_1}.

\begin{figure}
  \centering
\begin{tabular}{cc}
    \begin{minipage}{0.9\textwidth}
        \includegraphics[width = \textwidth]{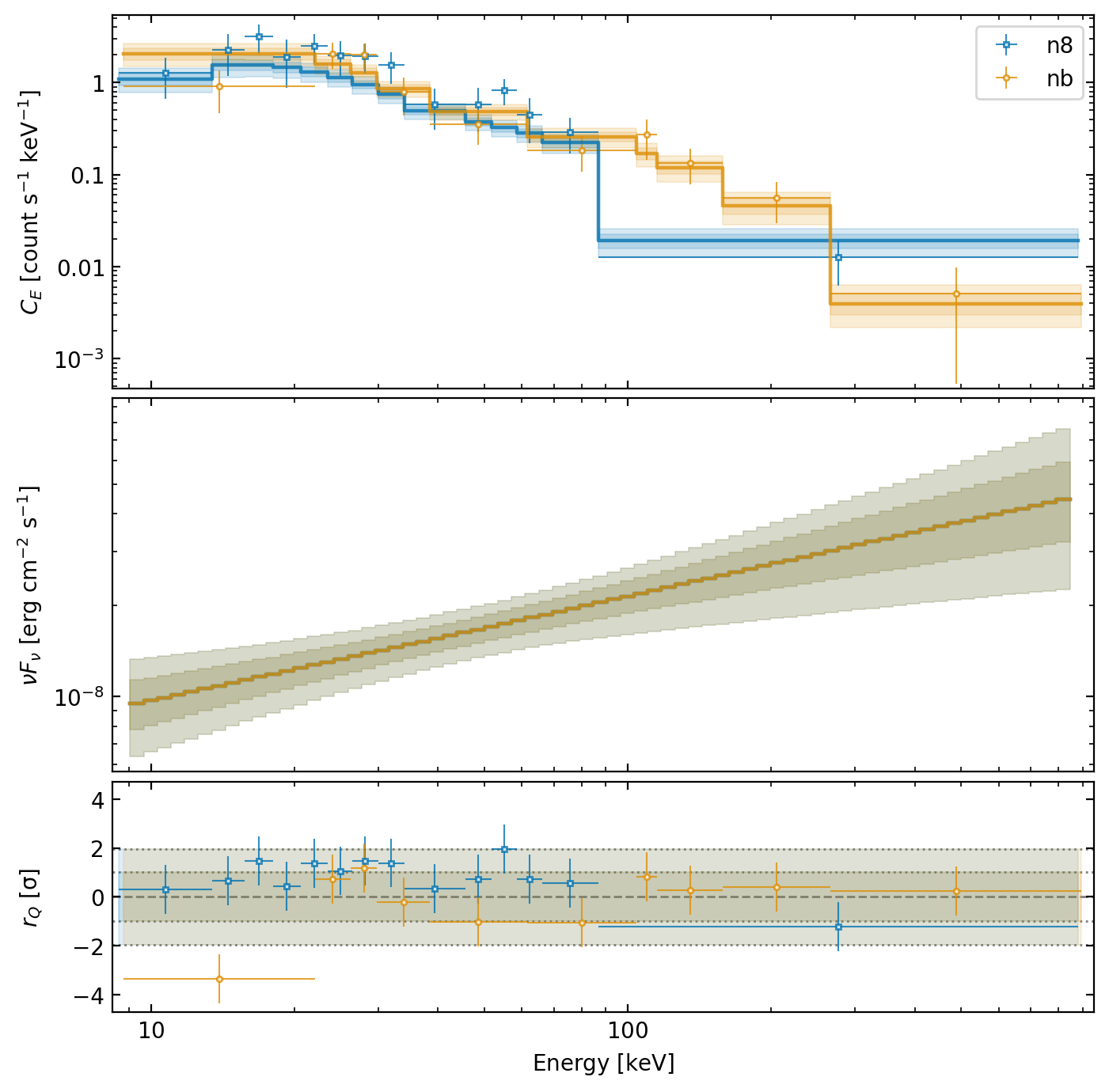}
    \end{minipage}\\
  \begin{minipage}{0.9\textwidth}
      \includegraphics[width = \textwidth]{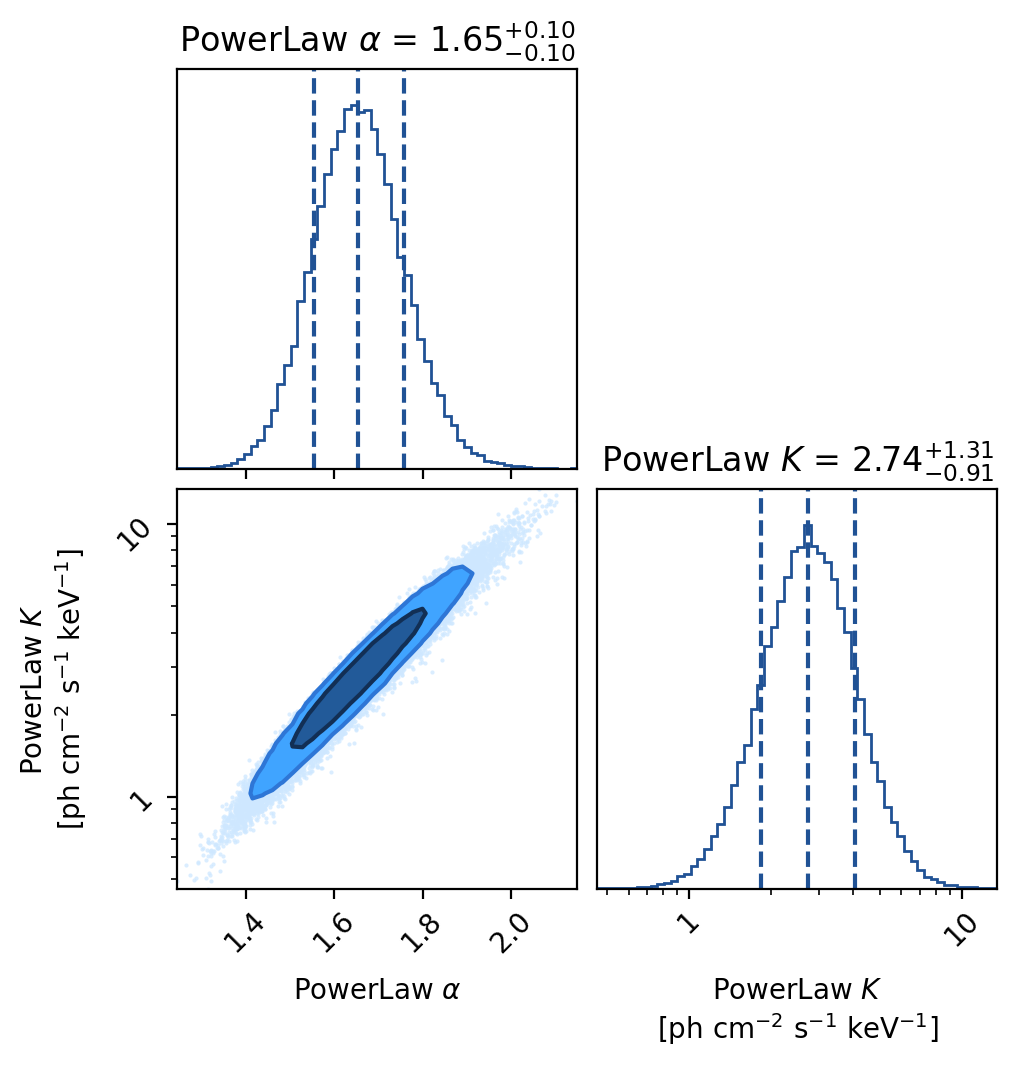}
   \end{minipage}
\end{tabular}
  \caption{\label{fig:pl_fit}\small The spectral fit result of the precursor candidate with power-law model.}
\end{figure}

\subsection{Bulk Periodicity} \label{sec:bulk_periodicity}

The multi-episode emission separated by quiescent intervals implies a possible periodicity behavior. We note that \cite{2025arXiv250714286L} claimed a possible periodicity of $\sim$ 2800\,s based on the light curve of \textit{Fermi}/GBM alone, however their analysis only considered a few specific pulses and ignored many other pulses in the burst.
Thanks to the more complete coverage provided by joint observation of \textit{Insight}-HXMT, GECAM-B and \textit{Fermi}/GBM, we can investigate the periodicity considering the full course of the main burst episodes as well as the precursor discovered in this work.

Since there are numerous weak spikes and pulses which are difficult to accurately identify, here we mainly estimate the period cycle based on the time intervals between those 4 large episodes of main burst,
also taking the durations and spiky structures into consideration 
(as marked in Fig.~\ref{fig:lc_overview}) for subsequent Monte-Carlo sampling method. When calculating the time intervals between different episodes, instead of applying the trigger time or pulse peak to represent each segment directly, we sample the time by taking the background-subtracted light curves as a probability distribution, obtaining a series of time points that represent the corresponding center of burst episodes. 
These sampled time points will be more concentrated around the pulse peak, but they are not fixed at a single moment. Especially when multiple pulses are present, this approach allows each pulse to potentially represent the corresponding explosion stage. 
The obtained time differences between adjacent episodes are 4200$^{+600}_{-600}$\,s, 3900$^{+2200}_{-1700}$\,s, 3800$^{+1700}_{-2200}$\,s with the mean value of 3900$^{+900}_{-900}$\,s.
As one can see, there is a potential quasi-periodicity.
This period of $\sim$ 4000\,s is also consistent with the Konus-Wind light curve which only covers the main burst episodes \citep{KW_GCN}. 
 Nevertheless, it is worth noting that the existence of periodicity as well as the period of $\sim$ 4000\,s are not very robust yet, mainly due to the limited cycle number and the uncertainties of period duration. Intriguingly, we find that
the precursor occurs about 23 cycles before the start time of episode M1.

\section{Soft X-ray emission} \label{section3}

We collect and analyze the public available soft X-ray observation data by Swift/XRT \citep{XRT_GCN} and the reported observation results from Chandra. Before the conclusion of its nature, we avoid to call it afterglow (with respect to the main burst). Since the soft X-ray emission follows the main burst in time, we call it ``following emission'' which is also denoted as episode F (see Fig. \ref{fig:lc_overview_all}) in this work.

To characterize the temporal evolution of these soft X-ray emission in episode F, we fit the X-ray flux data with the power-law model,
\begin{equation}
F \propto (t-T_0^{\rm PL})^k ,
\end{equation}
where $F$ is the flux in 0.3--10\,keV, $T_0^{\rm PL}$ is the start time of the power-law decay, and $k$ is the power-law index.
We set the start time ($T_0^{\rm PL}$) as free parameter and find the power-law fits the data well. Remarkably, the resulted power-law index is $-1.73^{+0.05}_{-0.04}$, which is almost perfectly consistent with the canonical -5/3 predicted for fallback in TDE or collapsar \citep{michel_neutron_1988,1989_phinney}. Also the start time of this power-law broadly aligns with the last episode of main burst (M4), which indicates a probable association between the gamma-ray emission in main burst (more specifically the episode M4) and soft X-ray emission in episode F. 

We also tried to fix the start time of the power-law decay to the time of the precursor, GRB 250702D, GRB 250702B and GRB 250702E, resulting in power-law index of -2.54, -1.94, 1.91, and -1.82, respectively. Indeed, a broken power-law (BPL), rather than a single power-law, is required to fit the data if setting the precursor as the start time,
\begin{equation}
F(t) = \begin{cases} 
F_\text{0}[(t-T_0^{\rm PL})/(t_\text{j}-T_0^{\rm PL})]^{-\alpha_1}, & t < t_\text{j} \\ 
F_\text{0}[(t-T_0^{\rm PL})/(t_\text{j}-T_0^{\rm PL})]^{-\alpha_2}, & t > t_\text{j} .
\end{cases}
\label{eq_bpl}
\end{equation}
The best fit result is shown in Fig. \ref{fig:lc_overview_all}. More details of these fittings can be found in Appendix \ref{sec:append_xray}.

We stress that, among these fittings, the one by setting the start time ($T_0^{\rm PL}$) as free parameter and resulting the canonical -5/3 power-law decay gives the best fitting statistics as shown in Table \ref{table_xrt_fit}, thus is the most probable scenario in our discussions below.
In addition, we note that it is usually desirable to fit the power-law evolution with a free start time rather than the trigger time of an instrument. An notable example is the discovery of the power-law evolution of the emission line from 37 MeV to 6 MeV in GRB 221009A \citep{zhang_observation_2024}, among which setting the initial time to the trigger time may also obtain a fit result but will lose the true physics.

\begin{figure}
    \centering
    \includegraphics[width=\linewidth]{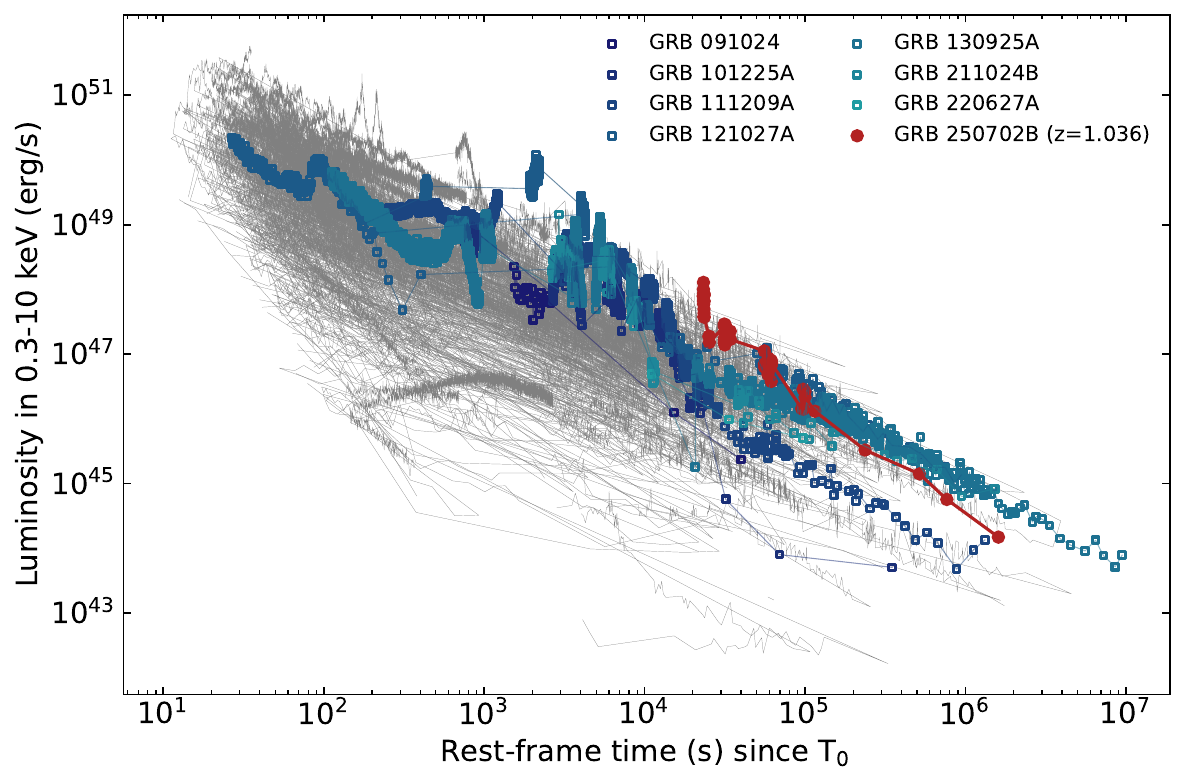}
    \caption{Soft X-ray emission of GRB 250702B/EP 250702a in comparison with GRBs including some ULGRBs. Redshift of GRB 250702B derived from \citet{gompertz_jwst_2025}}
    \label{fig:xray-ULGRBs}
\end{figure}

We compare the soft X-ray emission of this burst with other GRBs, especially ULGRBs, as shown in Fig. \ref{fig:xray-ULGRBs}. One can see that both the luminosity and temporal evolution of soft X-ray emission are generally consistent with known ULGRBs, which lie in the bright part of all GRBs. We suggest that this provides another evidence that 
this source is likely a ULGRB. 

\begin{deluxetable}{cccccccccc}
\tabletypesize{\scriptsize}
\setlength{\tabcolsep}{3pt}
\tablewidth{0pt}
\tablecaption{Fitting results for the soft X-ray data \label{table_xrt_fit}}
\tablehead{
 & & \multicolumn{3}{c}{$\text{power-law}$} & \multicolumn{5}{c}{\text{Broken power-law}}  \\
\cmidrule{3-5} \cmidrule(lr){6-10}
\colhead{Note} & \colhead{$T_0^{\rm PL}$(UTC)} & \colhead{k} & \colhead{$\rm BIC_{\text{PL}}$} & \colhead{$\rm AIC_{\text{PL}}$} & \colhead{$\alpha_1$} & \colhead{$\alpha_2$} & \colhead{$\Delta t_{\text{j}}^\textbf{a}$} & \colhead{$\rm BIC_{\text{BPL}}$} & \colhead{$\rm AIC_{\text{BPL}}$}\\
 \colhead{} & \colhead{} & \colhead{} & \colhead{} & \colhead{} & \colhead{} & \colhead{} & \colhead{(day)} & \colhead{} & \colhead{}
}
\startdata
Precursor & 2025-07-01T11:55:00 & $-2.54 \pm 0.03$ & 330.46 & 326.37 & $4.87 \pm 1.33$ & $1.96 \pm 0.13$ & $3.00 \pm 0.41$ & 45.36 & 37.19\\
$T_0$ & 2025-07-02T12:52:52 & $-1.95 \pm 0.02$ & 54.11 & 50.03 & -- & -- & -- & -- & --\\
GRB 250702D & 2025-07-02T13:09:02 & $-1.94 \pm 0.02$ & 50.97 & 46.88 & -- & -- & -- & -- & --\\
GRB 250702B & 2025-07-02T13:56:06 & $-1.91 \pm 0.02$ & 42.03 & 37.95 & -- & -- & -- & -- & --\\
GRB 250702E & 2025-07-02T16:21:33 & $-1.82 \pm 0.02$ & 17.99 & 13.91 & -- & -- & -- & -- & --\\
Free Fit & $\text{2025-07-02T16:04:19}^\textbf{b}$ & $-1.73 \pm 0.05$ & 8.87 & 2.74 & -- & -- & -- & -- & --\\
\enddata
\tablecomments{Parameter errors in this table are for 1 $\sigma$ (68\% confidence level). \textbf{a}:$\Delta t_{\text{j}} = t_{\text{j}} - t_0$. The unit of the  $\Delta t_{\text{j}}$ is the ``day". \textbf{b}: The 1 $\sigma$ error of $T_0^{\rm PL}$ is +3100\,s, -4700\,s for free-fit.
}
\end{deluxetable}

\section{Progenitor models}\label{section4}

First, we summarize the key observational properties of high energy (gamma-ray and soft X-ray) emission we found in GRB 250702B/EP250702a: Temporally, the gamma-ray emission of main burst last for about 4 hours. It is longer than 1 day if taking the precursor into account. The main burst could be grouped into 4 episodes, segmented by apparent quiescent periods. The waiting time between precursor and main burst is about 25 hours. The soft X-ray emission spans from -1 day to more than 30 days with respect to the main burst phase. The soft X-ray flux post-main burst follows a power-law decay of index -5/3 and the onset time of this power-law lies around the last episode of main burst. Not only the gamma-ray but also the following soft X-ray emission show high temporal variability. Spectrally, the precursor and the soft X-ray emission exhibit the same power-law index of about -1.6, which also agrees with the $\beta$ of Band spectrum for the main burst \citep{GBM_GCN_250702BDE}.

Next, we discuss the following some possible progenitor scenarios, primarily considering the observation facts summarized above.

\subsection{ULGRB: Self-regulated collapsar of supergiant star}

Firstly, we find that the luminosity and temporal evolution of soft X-ray emission resembles other ULGRBs, as shown in in Fig. \ref{fig:xray-ULGRBs} and discussed in Section \ref{section3}.

This burst is significantly longer than that of typical Type II GRBs -- the latter are widely thought to originate from the collapse of massive stars.
The duration of Type II GRB is generally believed to be determined by two key timescales: the stellar collapse timescale (e.g. typically tens of seconds to minutes for free-fall collapse of a Wolf-Rayet star) and the energy dissipation timescale (on the order of seconds to tens of seconds).
In the case of giant star, the inner layers likely collapse to form a black hole, launch relativistic jet and give rise to the prompt emission, while the outer layers will be blasted out. \cite{2012ApJ...752...32W} proposed that the collapse timescale of blue supergiant star could be extended to thousands of seconds, potentially explaining ULGRBs.
Depending on viewing angle, the observed emission duration can last up to $\sim 10^4$\,s, as shown in a simulation by \citet{perna_ultra-long_2018}.

If the progenitor is a more massive one, e.g. red supergiant star, taking Betelgeuse as an example, $R_B = 764 R_{\odot}$ and $M_B = 17.75 M_{\odot}$ (adopting the result of \citet{joyce_standing_2020}). Assuming homogeneous mass density, the free fall time scale of inner layer is
\begin{equation}
    t = \frac{\pi}{2} \sqrt{\frac{R_B^3}{2GM_B}} = 102.6 \text{\,day.}
\end{equation}
This time scale can serve as a lower limit ($\sim 100$\,day) of the fall back procedure.

Additionally, we note that, a red supergiant star can usually have dusty circumburst environment (e.g. SN2025pht with $A_V = 5.3\,\text{mag}$ reported by \citet{kilpatrick_type_2025}), thus can explain the optical extinction of GRB 250702B/EP250702a, $A_V = 11\,\text{mag}$, which is the sum of circumstellar and host galaxy dust extinction, as reported by \citet{2025arXiv250714286L}.

Regarding the high and fast variability in the observed soft X-ray emission of this burst, it is not expected by the afterglow thus we suggest it should originate from the internal region (prompt emission) of the jet which reflects the central engine activity. 
Remarkably, the $t^{-5/3}$ power-law decay of the X-ray flux
perfectly matches the theoretical prediction of free fall collapse with a constant density profile \citep{michel_neutron_1988}, indicating that the luminosity of the jet should be dominantly powered by the fallback accretion of the outer layers.

The fallback procedure of the outer layers of the supergiant star can last for long time, as the free fall timescale shown above. This can account for the observational fact from e.g. Swift/XRT and Chandra that -5/3 decay X-ray emission continues up to more than 60 days post burst \citep{eyles-ferris_grb_2025}. We note that similar fallback procedure has been used to explain X-ray bump in ULGRB afterglow \citep{wu_giant_2013}, and the light curve of a Type-II P supernova which could originate from a massive red supergiant \citep{wang_fallback_2018}.

The emission may decay to undetectable either because the emission is much weaker than the instrument sensitivity, or the jet may be shutdown when the late time accretion rate is too low.

Besides, in this fallback scenario, the onset time of the -5/3 power-law may correspond to the time when the outer layer is expelled by the violent activity of the central engine. The fact that the last episode (M4) of main burst occurs around the onset time the -5/3 power-law is very consistent with the aforementioned scenario.

Moreover, we suggest that the interplay between fall back accretion and strong outflow (and even the jet) in this scenario may drive quasi-periodic accretion across multiple epochs, which could naturally explain the observed intermittent bursts and possible bulk periodicity. This self-regulated collapse may also suppress the formation of a canonical core-collapse supernova (SN), providing an alternative explanation for the lack of SN counterpart of this burst yet.

\subsection{Jetted TDE: MS-IMBH or WD-IMBH}

In the TDE scenario, a relativistic jet is required to explain the highly variable and non-thermal emission in the gamma-ray and soft X-ray band. 
However, the existence of precursor and multi-episode main burst are not common in known jetted TDEs, neither the luminosity and temporal evolution of the soft X-ray (as suggested by \cite{2025arXiv250714286L}).

Remarkably, we find that the best fit of the X-ray flux data reveals that it follows a power-law decay with index of $\sim-5/3$ (when the start time of power-law $T_0^{\rm PL}$ is set as a free parameter). In the TDE model, this -5/3 decay is a canonical trend typically associated with the evolution of matter fallback rates \citep[e.g.][]{1988Natur.333..523R}. However, since the $T_0^{\rm PL}$ of the power-law decay corresponds to the moment of the tidal disruption of the star, the fact that the $T_0^{\rm PL}$ is about 28 hours after the precursor and about 3 hours after the emergence of the main burst in gamma-ray band raises a serious problem to interpret the precursor and main burst emission which precede the disruption of the star and mass fallback. If manually setting the $T_0^{\rm PL}$ to be precursor time or earlier time, the resulted power index will be steeper than $\sim-2.5$, which is just marginally allowed in TDE \citep[e.g.][]{gezari_tidal_2021}. Yet, the waiting time between the start time $T_0^{\rm PL}$ of the power-law decay and the first data that follows the -5/3 decay is only about 10 hours, which is again very unusual for TDE events, as the fall-back timescales typically observed in TDEs are usually in the order of days to months \citep{Sazonov_2021}.

Moreover, as noted in Section \ref{sec:bulk_periodicity}, although the limitations on the current data prevent us from drawing a robust conclusion on the periodicity, the burst light curves do show some sort of regular behavior with a bulk periodicity of about 4000 s.
If this periodicity is real, one might invoke the repeated partial tidal stripping of a WD by an IMBH to account for such period \citep[e.g.][]{eyles-ferris_can_2025}.
However, we argue that our results do not support such scenario.
As a degenerate stellar object, a WD exhibits an inverse correlation between mass and radius \citep{1935MNRAS..95..226C}: as a WD loses mass via partial stripping, the remaining core expands (increasing in radius) and becomes more vulnerable to tidal disruption in subsequent orbital passages. This would imply a progressive increase in the mass transferred to the IMBH with each orbit \citep{2023ApJ...947...32C} --- a trend that should be reflected in rising fluence across successive burst epochs. However, no such monotonic increase is observed in our data (see Figure~\ref{fig:lc_overview_all} and section~\ref{section2.2.1}). Additionally, this ``stripped-easier-to-strip" behavior of a partially stripped WD would likely lead to a runaway stripping even during a single periastron passage. As a result, the binary system would be unlikely to survive more than $\sim 4$ periastron passages. This problem will be more severe if considering the precursor about 23 period cycles before the main burst as well as the long-duration soft X-ray emission before the main burst. Consequently, we find that the observations presented in this work strongly disfavor the TDE scenario.

\subsection{Type IL GRB: Compact Merger}

Despite that the traditional theory of compact merger (e.g. NS-NS, BH-NS) can only lead to short duration GRB, it is worth to note that recent observations of some Type IL GRBs (e.g. GRB 230307A) indicate that the merger system can give rise to long duration burst \citep{2024Natur.626..737L, 2025ApJ...985..239Y, yi_long_2025}. Also Type IL GRBs are found to be composed of three stages of emission, i.e. precursor, main burst and extended emission \citep{TypeIL_1, TypeIL_2}, which seems to be somewhat consistent with the morphology of this burst. Indeed, how the compact merger of Type IL GRB can produce intrinsically long burst is not concluded yet. Additionally, we note that this burst has an energy fluence less but compatible to GRB 230307A, the typical Type IL GRB. Therefore, the possibility of merger system to produce ultra-long burst thus can account for this burst is worth to explore. Moreover, synergy operation of this event by O4 of LIGO-Virgo-KAGRA (LVK) provided us a great opportunity to test this idea. 

We checked the GW events reported by LVK during July 1st and 2nd, 2025. There are only two high-significance GW events, i.e. S250701bp and S250701bq, whose time and location do not conflict with this source. However, both of them are classified as BBH with more than 99\% probability, making them very unlikely to be associated with this burst. 

\subsection{Conclusion of models}

We note that,
there are many discussions and proposals on the progenitor of this burst. \citet{2025arXiv250714286L} discuss several progenitor scenarios and favor WD-IMBH scenario. 
Micro-TDE \citep{beniamini_ultra-long_2025, oconnor_comprehensive_2025} and helium-star merger \citep{neights_grb_2025} are also
proposed to explain this peculiar day-long GRB. 

However, as discussed above,
we find that the ULGRB model can adequately interpret 
the gamma-ray and soft X-ray observation facts we reported in this work. Especially, a self-regulated collapsar of a supergiant star can naturally unify the timescales and radiative properties: hour-scale inner collapse can sustain the ultra-long gamma-ray emission, days-to-year fallback of the outer layers can power the subsequent X-rays with a $t^{-5/3}$ power-law decay.
Such regulation may also suppress a canonical supernova, which is the case for some ULGRBs. Additionally, the circumburst environment of red supergiant star can account for the unusually large optical extinction of this burst.

\section{Discussions}\label{sec_discussions}

Under the ULGRB scenario, we discuss the more interpretations of the observed gamma-ray and soft X-ray emission of GRB 250702B/EP250702a.
Similar to other ULGRBs, the precursor and main burst could be readily interpreted by a relativistic jet launched during the collapse of a supergiant star. A two-step collapse \cite[e.g.][]{song_grb_2023} may be involved to account for the precursor and main burst: The first collapse may only launch a less energetic jet which produces a less luminous and shorter duration precursor than the main burst.
The main burst emission
are caused by a much more energetic jet launched during the second stage of collapse, which seems to last longer than the first stage. Moreover, the intermittent behavior (even with a quasi-periodicity) of the main burst may result from the interactions between the fallback material and the outflow.

The energetic explosion in the last episode of main burst (M4) probably triggers a further expansion --- and later fallback --- of outer layers of the supergiant star which gives rise to the post-main burst soft X-ray emission with a power-law decay index of -5/3, consistent with theoretically predicted value for free fall accretion \citep{michel_neutron_1988}.
Note that, the -5/3 power-law decay of soft X-ray emission deviates the temporal evolution of typical GRB afterglow, which is dominated by the interaction between jet and interstellar medium or progenitor wind. 

Moreover, the observation fact that the temporal variability in the soft X-ray \citep{swift_xrt_gcn} as well as the hard X-ray \citep{NUSTAR_GCN} during this power-law decay stage is much higher than the normal GRB afterglow also provides support to the above scenario.

Therefore, we conclude that 
both the gamma-ray emission in main episode and the following X-ray emission (Fig. \ref{fig:lc_overview_all}) are produced by the relativistic jet and controlled by the fallback rate during a collapse of supergiant star.

In the following, we try to infer the Lorentz factor of the jet during the main burst and the following soft X-ray episode. Due to limited observation and analysis at this stage, we only give an ``order of magnitude'' estimation.

First, we estimate the Lorentz factor of the main burst (denoted as $\Gamma_{\rm M}$)
using empirical relations.
Based on the ``Amati" relation \citep{Amati2002} and $E_{\rm \gamma,iso}$-$\Gamma_{0}$ relation \citep{2010Liang}, we find
\begin{equation}
\Gamma_{\rm M} \sim 133\left ( \frac{E_{\rm peak}}{100~\rm keV} \right )^{\frac{5}{8} }(1+z)^{\frac{5}{8} } .
\end{equation}

Using $E_{\rm peak}$ of about 400 ± 100\,keV of the main burst \citep{GBM_GCN_250702BDE} and the redshift ($z$) of 1.036 \citep{gompertz_jwst_2025}, we estimate the $\Gamma_{\rm M}$ of $410\sim570$. This estimation has a large uncertainty due to the dispersion of the relations and uncertainties on the $E_{\rm peak}$. It should be noted that the high energy photon index ($\beta$) of Band spectrum is about -1.6 \citep{GBM_GCN_250702BDE}, indicating that there is a break in higher energy range. If so, the $\Gamma_{\rm M}$ should be larger.
Next, we roughly estimate the Lorentz factor during the following soft X-ray episode with -5/3 power-law decay (episode F, about 1 to 30 days post main burst, denoted as $\Gamma_{\rm F}$). Given that the main burst and the following soft X-ray emission share the same power index, we assume that the spectrum in the jet co-moving frame maintains the same shape among these stages and the difference of observed spectrum is caused by the different Lorentz factor of the jet among these episodes, therefore we can roughly estimate the ratio of Lorentz factor between these two stages based on the lower energy boundaries of the energy band with spectral index of -1.6, $\frac{\Gamma_{\rm M}}{\Gamma_{\rm F}} \sim 100$. Thus, we can infer that $\Gamma_{\rm F}$ is about several (order of magnitude estimation), which corresponds to relatively mild accretion rate in late time fallback. 

\section{Summary and Conclusion}\label{section5}

In this work, we characterize the properties and investigate the nature of the peculiar ultra-long transient source GRB 250702B/EP250702a. We focus ourselves on the high energy emission, including gamma-ray and soft X-ray. Especially, we perform a comprehensive multiple-instrument burst search and implemented analyses jointly with Insight-HXMT/HE, GECAM-B and Fermi/GBM data. We also analyz the temporal evolution of soft X-ray emission from Swift/XRT and Chandra observations. Our findings are summarized as following:

1. In gamma-ray band, the Insight-HXMT and GECAM-B observation shows that the main burst starts much earlier than the Fermi/GBM trigger time. The main burst could be generally divided into four episodes (denoted M1 to M4), with a total duration of about 4 hours. Each episode consists of a bulk of numerous spiky pulses and some underlying slow components. These episodes are separated by quiescent periods. There is no monotonic increase or decrease in the fluence of these episodes, however there may be a possible (quasi-)periodicity of about 4000 s.

2. There is a gamma-ray precursor candidate (denoted as P) at 2025-07-01T11:55:17 UTC, which is about 25 hours before the main burst, with a duration of about 50 s. Discovery of this precursor is broadly consistent with the EP/WXT detection of this burst a day before the gamma-ray trigger of the main burst. The relation between the precursor-main burst waiting time and the duration of main burst is generally compatible to the trend of GRBs and favors a ULGRB.

3. The luminosity and temporal evolution of the soft X-ray emission resemble other ULGRBs very well, but apparently differs from the known jetted TDEs at the same stage.

4. The soft X-ray flux exhibits a striking feature of $t^{-5/3}$ power-law decay, which perfectly matches the canonical expectation of fallback accretion to a black hole which could exhibit in collapsar and TDE scenarios.
However, the fact that the initial time of the $t^{-5/3}$ power-law decay goes after the precursor and the onset of main burst in principle rejects the normal TDE scenario where all emission should be produced after the disruption of the star.
Moreover, the emergence of the -5/3 decay in the soft X-ray is only about 10 hours after the initial time of this power-law decay, which is also unusually short for normal TDEs.

5. These gamma-ray and soft X-ray emission share similarities of highly variable in flux and non-thermal spectrum with the similar spectral shape. They are probably produced by
a relativistic jet launched by a collapsar, whose luminosity and temporal evolution are dominated by the fallback process. Differences on the temporal and spectral behaviors of emission in different episode (P, M1 to M4, F) are likely primarily tracing the different Lorentz factor of the jet. The ultra-long duration of the fallback process requires a supergiant star as progenitor. 

In conclusion, all these observation features of GRB 250702B/EP250702a found in this work strongly support the ULGRB collapsar scenario. Almost all properties of precursor and main burst, as well as the luminosity and temporal evolution of soft X-ray emission, can be interpreted by a relativistic jet (with Lorentz factor varying from hundreds to several) launched during the collapse of a supergiant star.

Finally, we note that GRB 250702B/EP250702a represents the longest ULGRB ever detected, with a total gamma-ray duration of about 29 hours considering the precursor. In this event, we reveal a classic fallback process in the late stage of the collapse of a supergiant star which exhibits the $t^{-5/3}$ decay on the fallback rate.
Further multi-wavelength/multi-messenger observations and multi-instrument analyses are expected to uncover more mysteries of this peculiar transient.

\section*{Acknowledgments}
We appreciate the anonymous referee for helpful suggestions. This work is supported by 
the National Natural Science Foundation of China (Grant No. 12494572,
12273042
),
the National Key R\&D Program of China (2021YFA0718500),
the Strategic Priority Research Program, the Chinese Academy of Sciences (Grant No. 
XDB0550300
).
This work made use of data from the \textit{Insight}-HXMT mission, funded by the CNSA and CAS.
The GECAM (Huairou-1) mission is supported by the Strategic Priority Research Program on Space Science (Grant No. XDA15360000) of Chinese Academy of Sciences. 
We are grateful to the development and operation teams of \textit{Insight}-HXMT and GECAM.
We appreciate the public data and software of \textit{Fermi}/GBM. 
This work made use of data supplied by the UK Swift Science Data Centre at the University of Leicester.

\clearpage

\appendix

\renewcommand\thefigure{\Alph{section}\arabic{figure}}

\section{Evolution of soft X-ray emission} \label{sec:append_xray}

It is known that the different time zero will affect the slope of power-law evolution. 
To investigate this effect, as well as exploring genuine initial start time of decaying, we fit the soft X-ray emission, including observation from Swift/XRT and Chandra, by power-law model with different $T_0^{\rm PL}$.
The fitting results are summarized in Table~\ref{table_xrt_fit}, and the corresponding plots are shown in Fig. \ref{fig_xrt_fit}.

\setcounter{figure}{0}
\begin{figure}
  \centering
\begin{tabular}{c}
    \sidesubfloat[]{\includegraphics[width = .58\textwidth]{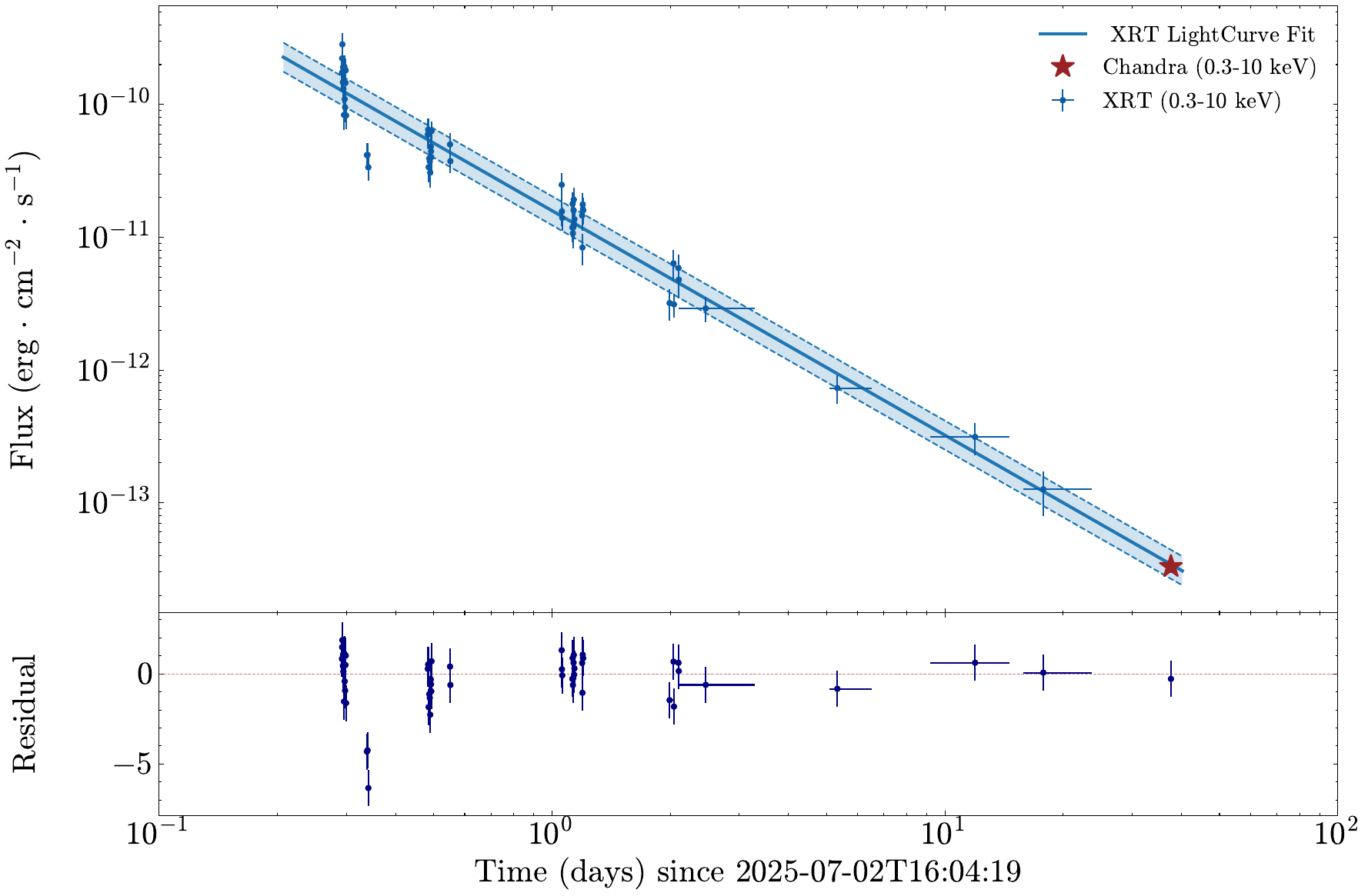}}
    \\
      \hspace{27pt}
    \sidesubfloat[]{\includegraphics[width = .6\textwidth]{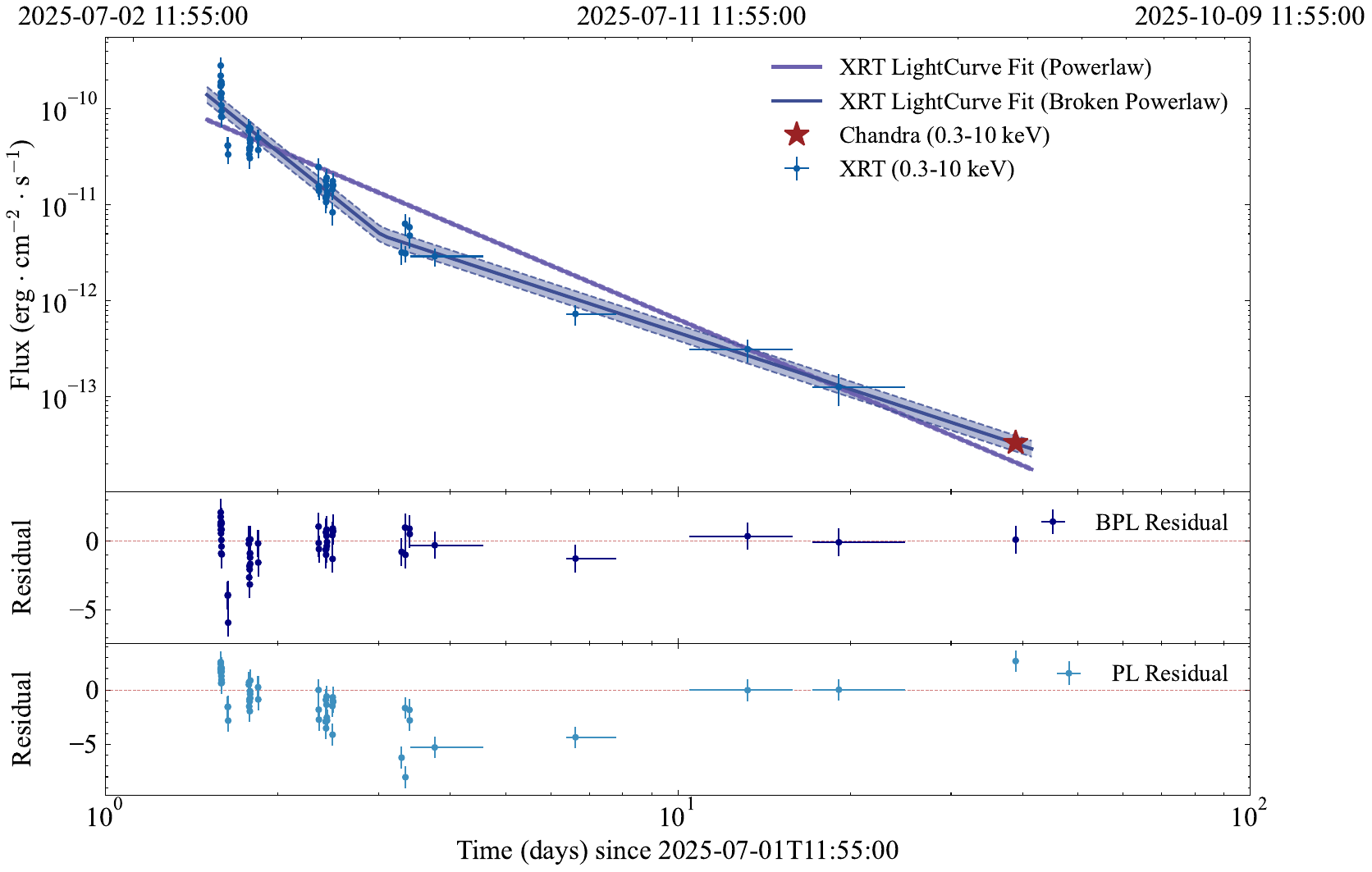}}
    \\
    \hspace{10pt}
    \sidesubfloat[]{\includegraphics[width = .6\textwidth]{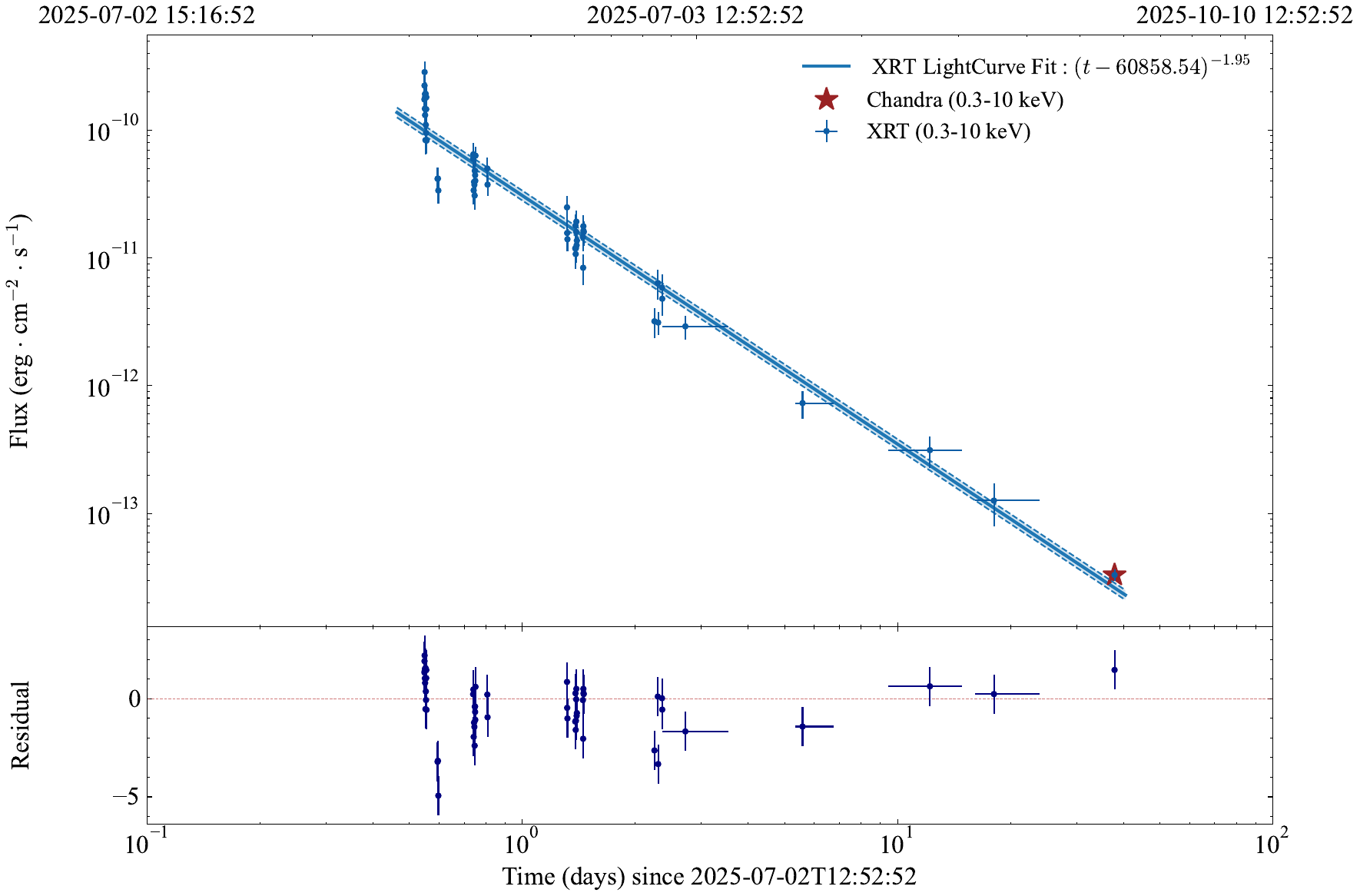}}
\end{tabular}
  \caption{\label{fig_xrt_fit}\small \textbf{(a),} the power-law model fitting curves and residual plots, whose $T_0^{\rm PL}$ is set as a free parameters duraing the fitting. \textbf{(b),} setting the trigger time of precursor as $T_0^{\rm PL}$, the fitted curves and residual for the BPL model and the power-law model. 
  \textbf{(c),} the power-law model fitting results with $T_0^{\rm PL}$, and the start time of the HXMT observation (i.e. 2025-07-02T12:52:52) is set as $T_0^{\rm PL}$.}
\end{figure}

Specifically, we note that, when setting the trigger time of precursor, which was detected about one day before the trigger of GRB 250702D, as the $T_0^{\rm PL}$, the power-law model provides a poor fit, and the trend in residuals suggests a possible BPL behavior, whose mathematical expression is expressed as Eq.~\ref{eq_bpl}. 
By comparing the Bayesian Information Criterion (BIC, defined as BIC$=-2\ln L+k\ln N$, where L represents the maximum likelihood value, $k$ denotes the number of free parameters in the model, and $N$ signifies the number of data points) and Akaike Information Criterion (AIC, defined as $-2\ln L+2k$) \citep{AIC} values obtained from fitting the power-law model and the BPL model, we find that both $\rm BIC_\text{PL} \gg BIC_\text{BPL}$ and $\rm AIC_\text{PL} \gg AIC_\text{BPL}$, indicating that the BPL model provides a much better fit than power-law model when the start time of the precursor is set as $T_0^{\rm PL}$, as shown in the Figure~\ref{fig_xrt_fit}(b). 
Additionally, the two slopes of the BPL model are $\sim$ -4.87 and $\sim$ -1.96, which are consistent with typical values of the steep decay phase and the decay phase after the jet break respectively, suggesting the soft X-ray emission can be explained as the afterglow is considered to have started from the precursor.

Additionally, we calculated the BIC and AIC for the power-law model with different $T_0^{\rm PL}$. The results show that as the $T_0^{\rm PL}$ approaches the value derived from the free-fit, the BIC and AIC values of the power-law model decrease, and the free-fit $T_0^{\rm PL}$ is the most favored value.

\section{Light curve and background analyses} \label{sec:append_lc_bg}

In order to reveal a complete day-long light curve, \textit{Insight}-HXMT/HE, GECAM-B and \textit{Fermi}/GBM are used to cover nearly full duration. The background of each instrument can change greatly in less than one hour, making the light curve from the source hard to recognize. Thus, background subtraction is necessary.

For \textit{Insight}-HXMT/HE, the background varies slowly when the satellite enters or quits the high-latitude area. Although most of the pulses of this burst detected by \textit{Insight}-HXMT/HE last for less than dozens of seconds, these pulses can be squished in a time interval of over a hundred seconds. In that case, 4th-order polynomial becomes less reliable for background fit.

We estimate the background of 400s to 1300s (part of M1) and 12100s to 12800s (part of M4) with revisited orbital data. For other time intervals, we use slide-window 4th-order polynomial to fit \textit{Insight}-HXMT/HE background.

Revisited orbital data background subtraction method utilized for \textit{Insight}-HXMT/HE is performed as follows:

First, we check satellite position and determine a 85500s raw shift period. We select number of shifted periods from -10 to 10 (except 0, which is the target lightcurve for background subtraction) and remove those with inconsistent satellite attitude. At least 3 effective orbits are selected at this stage. We find the lightcurves of these orbits with energy ranging from 200 keV to 3 MeV. Then, cross-correlation function (CCF) value between each pair of lightcurves is calculated for different refined shift period selection, ranging from 85480 to 85520 s. We use 85510 s as the best refined shift period, which maximizes the summed CCF value. The lightcurves at these revisited orbits are shifted according to refined shift period and the shift period number. The stacked lightcurve then can provide a better shape reference than 4-th order polynomial for the target lightcurve. Finally, we fit the background with the only one remained free parameter, the amplitude multiplier between the target lightcurve background and the stacked revisited lightcurve.

The background of GBM varies greatly and in many time intervals of this burst hard to fit with 4-th order polynomial.
For the background estimation of \textit{Fermi}/GBM, we employ the method using revisited orbital data, which has been adopted by \cite{09A_afterglow} and \cite{Biltzinger_2020}. After approximately 30 orbits, the Fermi/GBM returns to a nearby spatial location with a closely matched attitude, resulting in a nearly identical background environment for the GBM. The average background count rate from the ±30 orbit data (T ± 85060 s) of Fermi/GBM is used as an estimate of the background for the current orbit. This method remains applicable across different energy ranges, although some revisited orbits may exhibit particle contamination in the low-energy range. This contamination can significantly affect the background estimation. For instance, during the M2 period, the \textit{Fermi}/GBM background remains relatively stable, whereas particle event contamination has been observed in the revisited orbits immediately preceding and following this period. Fortunately, the background in the high-energy range remains relatively stable, and we assume that the spectral shape of the background remains constant during this interval. Thus, the variation of the background in the low-energy range can be extrapolated from its variation in the high-energy range. This procedure yields a background estimate that subsequently undergoes a straightforward angular correction, ultimately producing the net light curve for the GBM.

For HXMT/HE, all 18 HED are selected. For GECAM-B, 3 detectors with incident angle smaller than 70 $\deg$ are selected, namely GRD 12, 13, and 21. For \textit{Fermi}/GBM, to maintain favorable incident angles throughout all observational periods, the n8, na and nb detectors were selected for revisited orbital background subtraction.

Detailed light curves can be found in Fig. \ref{fig:lc_overview}.

\bibliography{reference}{}
\bibliographystyle{aasjournalv7}

\end{document}